\documentclass[journal]{IEEEtran}

\IEEEoverridecommandlockouts

\usepackage{caption}
\usepackage{subcaption}
\usepackage{graphicx}
\usepackage{cite}
\usepackage{url}
\usepackage[cmex10]{amsmath}
\usepackage{amssymb}
\usepackage{amsthm}
\usepackage{algorithm}
\usepackage{algorithmicx}
\usepackage{algpseudocode}
\usepackage{cases}
\usepackage{color}
\usepackage{soul}
\usepackage{xcolor}
\usepackage{framed}
\usepackage{cite}
\usepackage{epstopdf}
\usepackage{calc}
\usepackage{array}
\usepackage{amsmath}
\usepackage{comment}
\usepackage{bbm}
\usepackage{dsfont}
\colorlet{shadecolor}{yellow}
\DeclareGraphicsExtensions{.pdf,.eps,.png,.jpg,.mps}

\DeclareMathOperator{\diag}{diag}

\begin{document}
\title{Cyclic Weighted Centroid Algorithm for Transmitter Localization in the Presence of Interference}
\author{Shailesh Chaudhari and Danijela Cabric%
\thanks{Shailesh Chaudhari and Danijela Cabric are with the Department of Electrical Engineering, University of California, Los Angeles, 56-125B Engineering IV Building, Los Angeles, CA 90095-1594, USA (email: schaudhari@ucla.edu, danijela@ee.ucla.edu).}
\thanks{Part of this work has been published to the proceedings of IEEE GLOBECOM, 2014, Austin, Tx, USA~\cite{Shailesh2014}.}
\thanks{This material is based upon work supported by the National Science Foundation under Grant No. 1117600.}}
%
%
%


\maketitle

\begin{abstract}
This paper addresses the problem of localizing a non-cooperative transmitter in the presence of a spectrally overlapped interferer in a Cognitive Receiver (CR) network. It has been observed that the performance of non-cooperative Weighted Centroid Localization (WCL) algorithm degrades in the presence of a spectrally overlapped interferer. We propose Cyclic WCL algorithm that uses cyclic autocorrelation (CAC) of received signals at CRs in the network to estimate the location coordinates of the target transmitter. Performance of the proposed algorithm is further improved by eliminating CRs in the vicinity of the interferer from the localization process. In order to identify and eliminate CRs in the vicinity of the interferer, the ratio of the variance and the mean of the square of absolute value of the CAC, referred to as Feature Variation Coefficient (FVC), is used. Theoretical analysis of the Cyclic WCL algorithm is presented in order to compute the root mean square error in the location estimates. We  further study impacts of the interferer's power and location, CR density, and fading environment on the performance of Cyclic WCL. The comparison between Cyclic WCL and traditional WCL is also presented. 


\end{abstract}



\IEEEpeerreviewmaketitle



\section{Introduction}
\label{sec:Introduction}
The knowledge of transmitter location is required in  heterogeneous networks and cognitive radio networks for advanced spectrum sharing techniques including location-aware smart routing and location-based interference management \cite{Zhu2012}. In this paper, we address the problem of estimating the location coordinates of a non-cooperative transmitter (target) in the presence of a spectrally overlapped interference in a cognitive radio network. Under spectrally overlapped interference, the traditional Weighted Centroid Localization (WCL) algorithm \cite{Wang2011} results in higher localization errors, since it uses only the received signal power for localization. In this work, we utilize distinct cyclostationary features of the overlapping signals that arise from different symbol rates, in order to estimate the location of the target transmitter. We propose Cyclic WCL and improved Cyclic WCL algorithms to estimate the target location in the presence of spectrally overlapped interference.

Spectrally overlapped interference arises in the following two scenarios. First, let us consider a scenario where the target transmitter (Tx-1) and the interferer (Tx-2) belong to two different networks sharing a common frequency spectrum in a heterogeneous network. For example, consider that Tx-1 is a Wi-Fi transmitter, while Tx-2 is an LTE base station coexisting in the same frequency band. Such coexistence of LTE and Wi-Fi networks is considered for LTE-unlicensed band around 5GHz \cite{Zhang2015,Sagari2015,Bhorkar2014,Cano2015,Paiva2013,Nihtila2013}. The medium access protocol of the two networks is different and the lack of coordination among the two networks results in interfering transmissions of Tx-1 and Tx-2. In this case, the cyclostationary features of the signals transmitted from Tx-1 and Tx-2 are different due to different symbol rates in LTE and Wi-Fi systems. The cognitive receivers (CRs) in the network can exploit the distinct cyclostationary properties of Tx-1 in order to estimate its location coordinates. 
	
For the second scenario,  we consider transmitter localization problem in a CR network in the presence of a jammer \cite{Camilo2012, Fragkiadakis2013, Hlavacek2014} that transmits energy in the same band as Tx-1. In this case, the traditional WCL algorithm \cite{Wang2011} results in higher localization error because it uses only the received signal power at each CR. On the other hand, the proposed Cyclic WCL and improved Cyclic WCL provide robustness against such interference by using distinct cyclostationary properties of the Tx-1 signal.

Non-cooperative localization is commonly used in CR networks where the target transmitter does not cooperate with CRs in the localization process. In this scenario, localization techniques based on Time of Arrival (ToA) or Time-Delay of Arrival (TDoA) are not applicable. In this paper, we consider techniques that do not require ToA or TDoA information and estimates the target location using the received signal at the CRs in the network. 

\subsection{Related Work}
Several non-cooperative localization techniques have been proposed in the literature to estimate the location coordinates of the target transmitter. These techniques can be broadly classified as range-based \cite{Langendoen2003, Xiao2007, Chen2010} and range-free 
\cite{Bulusu2000, Blumenthal2007,  Laurendeau2010,  Ma2010,  Wang2011, Mariani2012, Xu2011, Zhang2012, Nanda2012, Kong2013, Feng2013, Zhao2014}. Range-based techniques  provide better estimate of the target location than range-free techniques, but require accurate knowledge of  wireless propagation properties such as path-loss exponent. Accurate information of the path-loss exponent is difficult to obtain and it may not be available. In such a case, range-free techniques such as centroid localization \cite{Bulusu2000, Blumenthal2007, Laurendeau2010} are used to perform coarse-grained target localization without any knowledge of the path-loss exponent. 

In the Weighted Centroid Localization (WCL) algorithm \cite{Blumenthal2007, Laurendeau2010}, the target location is approximated as the weighted average of all CR locations in the network. The CR locations are known at the central processing node where the WCL algorithm is implemented.  There are different variations of the WCL algorithm proposed in the literature to improve the localization performance \cite{Wang2011, Mariani2012, Xu2011, Zhang2012, Nanda2012, Kong2013, Feng2013, Zhao2014}. Theoretical analysis of the WCL algorithm and its distributed implementation are presented in \cite{Wang2011}. Different weighting strategies and CR selection techniques are presented in \cite{Mariani2012} to reduce the adverse impact of the border effect and to improve root mean square error (RMSE) performance of the algorithm. The technique presented in \cite{Xu2011} selects reliable CRs that are located closer to the target and uses them in the localization process. Papers \cite{Zhang2012} and \cite{Nanda2012} propose direction vector hop (DV-hop) and received signal power based fuzzy logic interference model, respectively, to assign WCL weights. The WCL algorithm that takes into account self-localization error of CRs is presented in \cite{Zhao2014}.

In the algorithms presented in \cite{Blumenthal2007, Laurendeau2010, Wang2011, Mariani2012, Xu2011, Zhang2012, Nanda2012, Kong2013, Feng2013, Zhao2014}, the weights for each CR location are computed based on the received signal power from the target. However, in the presence of a spectrally overlapped interference in the network, the received signal power at each CR is the summation of powers received from the target and the interferer. Therefore, localization errors in existing algorithms increase significantly due to the spectrally overlapped interference. Hence, there is a need to modify the WCL algorithm when a spectrally overlapped interferer is present in the network. 

\subsection{Summary of Contributions and Outline}
In this paper, we propose Cyclic WCL algorithm to estimate the location coordinates of the target transmitter (Tx-1) in a CR network in the presence of a spectrally overlapped interference (Tx-2). The location coordinates of the CRs in the network are known at the central processing node, where the localization algorithm is implemented. The cyclostationary properties of the target signal are used to compute weights in the Cyclic WCL algorithm. The proposed localization algorithm does not require any knowledge of the path-loss model and the location of the interferer.

The contributions of this paper are as follows:
\begin{enumerate}
\item
A cyclostationarity-based localization algorithm referred to as Cyclic WCL is proposed in order to estimate the target location in the presence of a spectrally overlapped interference. In the proposed algorithm, the central processing node estimates the target location as a weighted sum of the CR locations, where weights for CR locations are computed based on the cyclic autocorrelation (CAC) of the received signal at that CR.

\item
Theoretical analysis of the proposed algorithm is presented. The RMSE in the location estimate is computed as a function of the target and the interferer locations, their transmitted powers, and the CR locations.

\item
The performance of Cyclic WCL is further improved by eliminating CRs in the vicinity of the interferer from the localization process. The ratio of the variance and mean of the square of absolute value of the CAC of the received signal at each CR is used to identify and eliminate CRs in the vicinity of the interferer. Theoretical analysis for the improved Cyclic WCL is also presented in order to compute the RMSE.
\end{enumerate}

This paper is organized as follows. Section \ref{sec:Model} describes the system model, Cyclic WCL and improved Cyclic WCL algorithms. Section \ref{sec:Theoretical} presents theoretical analysis to compute the RMSE. Simulation results in various scenarios are provided in Section \ref{sec:Results}. Finally, Section \ref{sec:Conclusion} concludes this work. 

In this paper, we denote vectors by bold, lowercase letters, e.g., $\mathbf{a}$ or $\boldsymbol{\theta}$. Matrices are denoted by bold, uppercase letters, e.g., $\mathbf{A}$. Scalars are denoted by non-bold letters, e.g., $x_k$ or $R_{r_k}$. Transpose, conjugate, trace and determinant of matrices are denoted by $(.)^T$, $(.)^*$, $\operatorname{Tr}(.)$, and $\operatorname{det}(.)$, respectively. Finally, the terms Tx-1 and target are used interchangeably. 

\section{System Model and Algorithm}
\label{sec:Model}
\subsection{System Model}
\begin{figure*}
\centering
\includegraphics[width=\columnwidth]{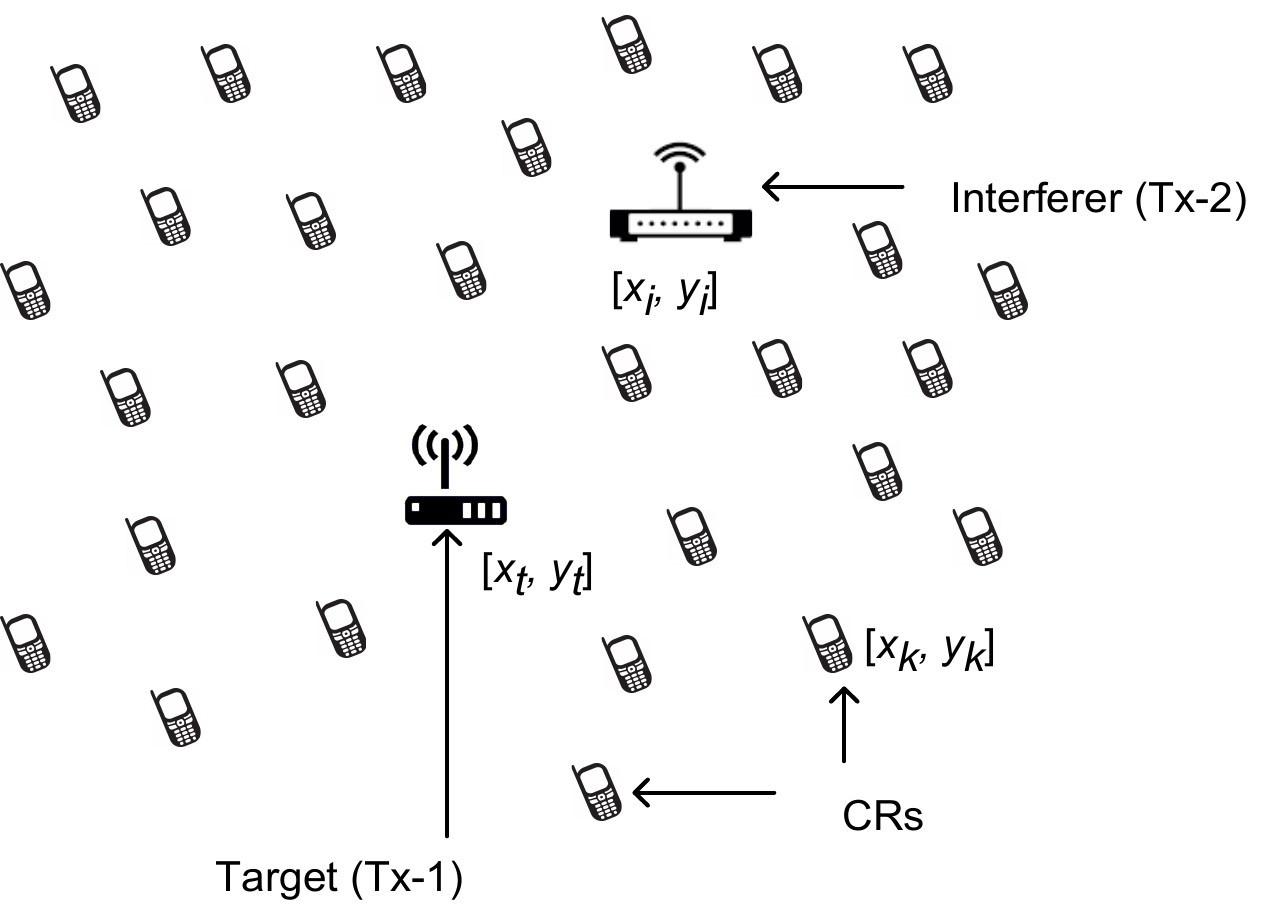}
\caption{System schematic: CR network with a target (Tx-1) and a spectrally overlapped interferer (Tx-2). The location coordinates of the target and the interferer are [$x_t, y_t$] and [$x_i, y_i$] respectively. The CRs are located at [$x_k, y_k$], $k = 1,2...K$.}
\label{fig:system_schematic}
\end{figure*}
In this paper, we consider a target-centric CR network, where the target is located at the origin and its coordinates are given by $\mathbf{L_{t}}=[x_{t},y_{t}]^{T}=[0,0]^{T}$. There are $K$ CRs in the network, distributed in a square of side $a$ meters around the origin. The locations of the CRs are known at the central processing node and are denoted by $\mathbf{L_{k}}=[x_{k},y_{k}]^{T}, 1\leq k\leq K$. The interferer is located at $\mathbf{L_{i}}=[x_{i},y_{i}]^{T}$. The knowledge of the interferer's location is not required while estimating the target's location coordinates. We assume that the target and the interferer locations are different, i.e., $\mathbf{L_t} \neq \mathbf{L_i} $. The schematic of the system is shown in Fig.~\ref{fig:system_schematic}.

Let $p_t$ and $p_i$ denote transmitted powers from the target and the interferer, respectively. The signals transmitted from the target and the interferer are $\sqrt{p_t}s_t(t)$ and $\sqrt{p_i}s_i(t)$, respectively. Both $s_t$ and $s_i$ are unit power signals. The signals $s_t$ and $s_i$ partially or completely overlap in the frequency domain. 
The algorithm presented in this paper is applicable to single carrier as well as multi-carrier OFDM signals, since both signal types exhibit cyclostationary properties. The single carrier and multi-carrier signal models are described below.
\subsubsection{Single carrier signal model}
\label{sec:single_carrier_model}
Single carrier target and interference signals can be expressed as
\begin{align}
\nonumber
s_{t}(n)=s_{t}(nT_{s})&=\sum_{l=-\infty}^{\infty}a_{l}g(nT_{s}-lT_{g})e^{j2\pi f_tnT_s}\\
&=\sum_{l=-\infty}^{\infty}a_{l}g_{n,l}e^{j2\pi f_tnT_s} \text{ and}
\label{eq:single_carrier_signal_model}
\end{align}
\begin{align}
\nonumber
s_{i}(n)=s_{i}(nT_{s})&=\sum_{l=-\infty}^{\infty}b_{l}h(nT_{s}-lT_{h})e^{j2\pi f_inT_s}\\
&=\sum_{l=-\infty}^{\infty}b_{l}h_{n,l}e^{j2\pi f_inT_s},
\label{eq:stsi}
\end{align}
where $T_{s}$ is the sampling period, $a_{l}$ and $b_{l}$ are transmitted
data symbols, $g$ and $h$ are pulse shaping filters, and $T_{g}$ and
$T_{h}$ are symbol periods of the signals $s_t$ and $s_i$, respectively. The carrier frequencies of $s_t$ and $s_i$ are $f_t$ and $f_i$, respectively. The data symbols $a_{l}$ and $b_{l}$ are assumed to be i.i.d. and zero mean. For simplicity of notation, we write $g_{n,l} = g(nT_s -lT_g)$ and $h_{n,l} = h(nT_s -lT_h)$. 

\subsubsection{Multi-carrier signal model}
\label{sec:multi_carrier_model}
An OFDM target signal with $N_{c,t}$ sub-carriers and sub-carrier spacing $\Delta f_{t}$ can be expressed as \cite{Sutton2008}
\begin{align}
\nonumber s_t(n) &= \sum_{l=-\infty}^{\infty}\sum_{\kappa=-\frac{N_{c,t}}{2}}^{\frac{N_{c,t}}{2}-1}c_{\kappa,l}e^{j2\pi (f_t + \kappa\Delta f_{t} )n T_s}g(nT_s - l T_g)\\
& = \sum_{l=-\infty}^{\infty}\sum_{\kappa=-\frac{N_{c,t}}{2}}^{\frac{N_{c,t}}{2}-1}c_{\kappa,l} g_{n,l} e^{j2\pi \kappa \Delta f_{t} n T_s} e^{j2\pi f_t n T_s},
\label{eq:ofdm_signal_model}
\end{align}
where $c_{\kappa,l}$ are data symbols on $\kappa^{th}$ sub-carrier, $g_{n,l}=g(nT_s -lT_g)$ is the window function. The duration of OFDM symbol is  $T_g = 1/\Delta f_t + T_{cp}$, where $T_{cp}$ is the duration of cyclic prefix. The data symbols $c_{\kappa,l}$ are assumed to be i.i.d. and zero mean. 
 Similarly, a multi-carrier interfering signal can be expressed as
\begin{align}
\nonumber s_i(n) &= \sum_{l=-\infty}^{\infty}\sum_{\kappa=-\frac{N_{c,i}}{2}}^{\frac{N_{c,i}}{2}-1}d_{\kappa,l}e^{j2\pi (f_i + \kappa\Delta f_{i} )n T_s}h(nT_s - l T_h)\\
& = \sum_{l=-\infty}^{\infty}\sum_{\kappa=-\frac{N_{c,i}}{2}}^{\frac{N_{c,i}}{2}-1}d_{\kappa,l}h_{n,l} e^{j2\pi \kappa \Delta f_{i} n T_s} e^{j2\pi f_i n T_s},
\end{align}
where $N_{c,i}, d_{\kappa,l}$, and $h_{n,l}$ denote the number of sub-carriers, symbols on $\kappa^{th}$ sub-carrier, and the window function, respectively.

Let $\alpha_t$ and $\alpha_i$ be the cyclic frequencies of the target and the interferer signal, respectively. For single carrier signals, the values of the cyclic frequencies of a signal depend on its modulation type and symbol rate \cite{Rebeiz2013}. For the CAC used in this paper, the cyclic frequencies are at integer multiples of the symbol rate. It should be noted that OFDM signals also have cyclic frequencies at integer multiples of symbol rate due to the existence of cyclic prefix\cite{Sutton2008}. 

In this paper, we assume that the cyclic frequencies of the target and interferer signals are different, i.e., $\alpha_i \neq \alpha_t$. The cyclic frequency of the target signal is known at the CRs in the network. To simplify the analysis, we consider sampling frequency $f_s=1/T_s > \text{max}(\alpha_t, \alpha_i)$. 

Let us denote the powers received at the $k^{th}$ CR from the target and the interferer by $p_{t,k}$ and $p_{i,k}$, respectively ($p_{t,k}^{\text{dB}}$ and $p_{i,k}^{\text{dB}}$ in logarithmic scale). In order to compute  the received powers, we take into account path-loss and shadowing effects. We assume that the signals $s_t$ and $s_i$ undergo independent shadowing due to the location difference in the target and the interferer. Therefore, $p_{t,k}^{\text{dB}}$ and  $p_{i,k}^{\text{dB}}$ are given by
\begin{align}
\nonumber  {p_{t,k}^{\text{dB}}} &= {p_t^{\text{dB}}} - 10\gamma \log \left(\frac{{||\mathbf{{L_k} - {L_t}}||}}
{{{d_0}}}\right ) - {q_{t,k}} \text{ and}\\
  {p_{i,k}^{\text{dB}}} &= {p_i^{\text{dB}}} - 10\gamma \log \left(\frac{{||\mathbf{{L_k} - {L_i}}||}}
{{{d_0}}} \right) - {q_{i,k}},
\label{eq:p_tk_p_ik}
\end{align}
where $p_t^\text{dB}$ is the power transmitted (in dB) by the target and $p_i^\text{dB}$ is the power transmitted by the interferer, $\gamma$ is the path-loss exponent and $d_0$ is reference distance. The variables $q_{t,k}$ and $q_{i,k}$ are used to model the shadowing effect on the target and the interferer powers. As mentioned above, $q_{t,k}$ and $q_{i,k}$ are independent variables and are not identical due to different locations of the target and the interferer. 

Let us consider a vector $\mathbf{q_t} = [q_{t,1},q_{t,2},...q_{t,K}]^T$ consisting of shadowing variables for the power received from the target at $K$ CRs. We consider uncorrelated shadowing at $K$ CRs in the network. For log-normal shadowing effect, $\mathbf{q_t}$ is modeled as a Gaussian random vector with zero mean and covariance matrix  $\sigma_q^2\mathbf{I_K}$, where $\mathbf{I_K}$ is a $K$x$K$ identity matrix ($\mathbf{q_t}\sim N(0,\sigma_q^2\mathbf{I_K})$). We assume that the shadowing statistics for the target and the interferer power are the same. Therefore, the vector corresponding to the interferer power is $\mathbf{q_i} =  [q_{i,1},q_{i,2},...q_{i,K}]^T \sim N(0,\sigma_q^2\mathbf{I_K})$. 

Further, the noise power at each CR is denoted by $\sigma_w^2$. The AWGN noise at $k^{th}$ CR is $w_k\sim CN(0,\sigma^2_w)$. Each CR samples the received signal at the sampling frequency $f_s=1/T_s$. The received signal at $k^{th}$ CR is given as
\begin{equation}
r_k(nT_s)=r_k(n)=\sqrt{p_{t,k}} s_t(n)+\sqrt{p_{i,k}} s_i(n)+ w_k(n).
\label{eq:rk}
\end{equation}

\subsection{Cyclic WCL}
\label{Cyclic WCL}
In Cyclic WCL, the CRs do not require the knowledge of transmitted powers ($p_{t}, p_{i}$) or received powers ($p_{t,k}, p_{i,k}$) in order to estimate the target location. Each CR in the network observes $N$ samples of the received signal $r_k$ and computes the non-asymptotic CAC of $r_k$ at the known cyclic frequency $\alpha_t=1/T_g$ of the target signal using
\begin{equation}
{\hat R_{{r_k}}^{\alpha_t}} = \frac{1}
{N}\mathop \sum \limits_{n = 0}^{N-1} |{r_k}(n){|^2}{e^{ - j2\pi {\alpha _t}n{T_s}}}.
\label{eq:cac_rk}
\end{equation}
In the above equation, (\^{}) indicates non-asymptotic estimate based on $N$ samples. In order to reduce the impact of the interferer signal at the cyclic frequency $\alpha_t$, the number of samples $N > 10 \lceil \frac{f_s}{\Delta\alpha}\rceil$, as shown in Appendix C, where $\Delta \alpha = |\alpha_t - \alpha_i|$ and $\lceil x\rceil$ denotes the smallest integer not less than $x$. In order to find $N$, the knowledge of $\Delta \alpha$ is required. In our system, $\alpha_t$ is known and $\alpha_i$ can be estimated using cyclostationary spectrum sensing \cite{Rebeiz2013}. From the estimate of $\alpha_i$, $\Delta\alpha$ and hence $N$ can be estimated. It should be noted that the knowledge of $\alpha_i$ is required only to obtain a lower bound on $N$ and the proposed algorithm does not depend on the accuracy of estimation of $\alpha_i$, as shown in results in Section \ref{sec:Results_Delta_Alpha}.

Using $\hat{R_{r_k}^{\alpha_t}}$ from (\ref{eq:cac_rk}), the central processing node estimates the location coordinates of the target as
\begin{equation}
\mathbf{{\hat L}_t} = \frac{{\sum\limits_{k = 1}^K {|{\hat R_{{r_k}}^{\alpha_t}}{|^2}} \mathbf{L_k}}}
{{\sum\limits_{k = 1}^K {|{\hat R_{{r_k}}^{\alpha_t}}{|^2}} }}.
\label{eq:lt}
\end{equation}
It should be noted that the locations of the CRs, the target and the interferer should remain constant until $N$ samples of the received signal are collected. Therefore, we assume that the locations do not change for time duration $NT_s$. We also assume that received powers $p_{t,k}$ and $p_{i,k}$ remain constant for this duration.

\begin{algorithm}[t]
\caption{Cyclic WCL}
\label{alg:cyclic_wcl}
\begin{algorithmic}[1]
\State Estimate $\Delta\alpha = |\alpha_t - \alpha_i|$ and select $N > 10 \lceil \frac{f_s}{\Delta\alpha}\rceil$.
\State At each CR $k=1,2..K$, collect $N$ samples of the received signal $r_k(n), n=1,2..N$.
\State At each CR, compute one CAC estimate: 
\Statex ${\hat R_{{r_k}}^{\alpha_t}} = \frac{1}
{N}\mathop \sum \limits_{n = 0}^{N-1} |{r_k}(n){|^2}{e^{ - j2\pi {\alpha _t}n{T_s}}}$
\State Compute the target location estimate at the central processing node:
$\mathbf{{\hat L}_t} = \frac{{\sum\limits_{k = 1}^K {|{\hat R_{{r_k}}^{\alpha_t}}{|^2}} \mathbf{L_k}}}
{{\sum\limits_{k = 1}^K {|{\hat R_{{r_k}}^{\alpha_t}}{|^2}} }}$
\end{algorithmic}
\end{algorithm}

\subsection{Improved Cyclic WCL}
\label{sec:Improved Cyclic WCL}
From the expression of the Cyclic WCL estimates in (\ref{eq:lt}), it is observed that the target location estimates are the weighted average of the CR locations and the weights are computed using the strength (square of absolute value) of the CAC of the received signal at each CR. Therefore, the target location estimate $\mathbf{{\hat L}_t}$ is closer to the CR which has the highest value of $\hat R_{r_k}$. From (\ref{eq:rk}) and (\ref{eq:lt}), intuitively it is expected that the location estimates are closer to the target's actual location when the interferer power is low. As the interferer power increases, the value of $\hat{R}_{r_k}$ at CRs in the vicinity of the interferer increases. Therefore, the location estimates move away from the target and towards the interferer's location. However, by eliminating the CRs in the vicinity of the interferer from the localization process, the impact of the interferer can be reduced. Based on this argument, we propose the improved Cyclic WCL algorithm that reduces the location estimation error due to the interferer by eliminating the CRs in the vicinity of the interferer.

First, let us define the transmit power ratio ($\rho$) and the received power ratio at the $k^{th}$ CR ($\rho_k$) as
\begin{align}
\rho = \frac{p_t}{p_i} \text{, and }
\rho_k = \frac{p_{t,k}}{p_{i,k}}. 
\end{align}
In the above equation, it is observed that if the $k^{th}$ CR is in the proximity of the interferer, we have $p_{i,k} > p_{t,k}$. On the other hand, for a CR in the proximity of the target, we have $p_{i,k} < p_{t,k}$. Therefore, the received power ratio ($\rho_k$) for a CR in the proximity of the interferer is smaller than for a CR in the proximity of the target. Further, we define Feature Variation Coefficient (FVC) at the $k^{th}$ CR as
\begin{equation}
\phi_k = \frac{\operatorname{var}(\hat R_{r_k}^{\alpha_t})}{E[|\hat R_{r_k}^{\alpha_t}|^2]}.
\label{eq:phi_k}
\end{equation}
As shown in the Appendix C and D, for a sufficiently large value of $N$ $\left(N > 10 \lceil \frac{f_s}{\Delta\alpha}\rceil \right)$, $\phi_k$ is a strictly monotonically decreasing function of $\rho_k$ and $0 \leq \phi_k \leq 1$ and it can be written as
\begin{align}
{\phi _k} &= \frac{{\operatorname{var} ({{\hat R}_{{r_k}}^{\alpha_t}})}}
{{E[|{{\hat R}_{{r_k}}^{\alpha_t}}{|^2}]}} = \frac{\rho_k^2\operatorname{var} ({{\hat R}_{{s_t}}^{\alpha_t}}) + E[|{{\hat R}_{{s_i}}^{\alpha_t}}{|^2}] + \rho_k E[|{{\hat R}_{{s_t}{s_i}}^{\alpha_t}}{|^2}]}
{\rho_k^2E[|{{\hat R}_{{s_t}}^{\alpha_t}}{|^2}] + E[|{{\hat R}_{{s_i}}^{\alpha_t}}{|^2}] + \rho_k E[|{{\hat R}_{{s_t}{s_i}}^{\alpha_t}}{|^2}]}.
\end{align}

\begin{figure}
\centering
\includegraphics[width= \columnwidth]{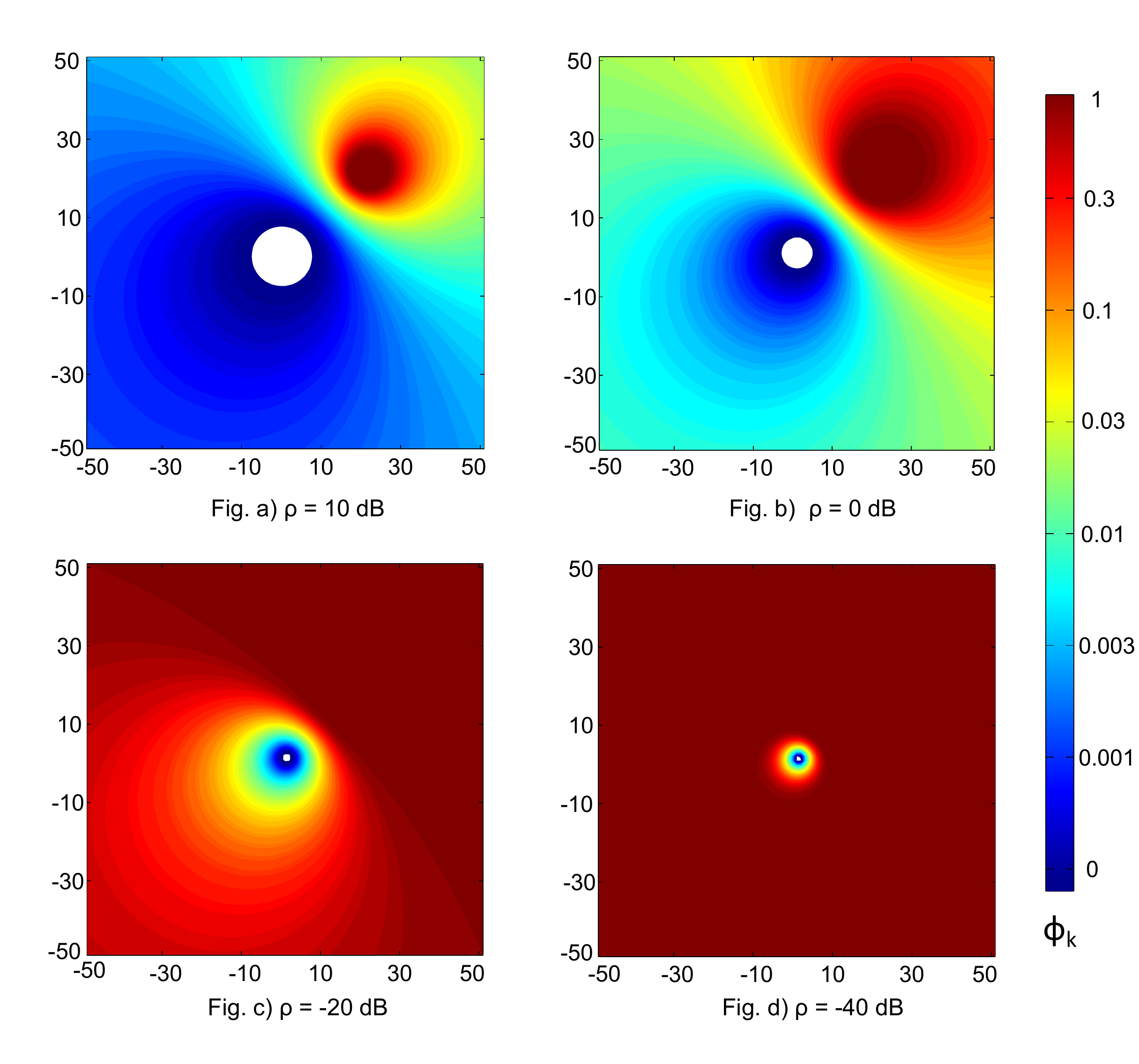}
\caption{Contour of FVC ($\phi_k$) at different locations in the network for a particular transmit power ratio ($\rho$). Target is at [0, 0]. Interferer is at [20, 20]. The target signal $s_t$ is 4-QAM with BW = 20MHz and carrier frequency = 2.4GHz. The interferer signal $s_t$ is 4-QAM with BW = 25MHz and carrier frequency = 2.4GHz. Shadowing variance $\sigma_q = 0$dB.}
\label{fig:fvc_contour}
\end{figure}

In order to illustrate how $\phi_k$ depends on the location of the $k^{th}$ CR, contour plots of $\phi_k$ at various locations in the network are shown in Fig. \ref{fig:fvc_contour}. It is observed that in the proximity of the target, $\phi_k \rightarrow 0$ . On the other hand, $\phi_k \rightarrow 1$ in the proximity of the interferer. In the four figures in Fig. \ref{fig:fvc_contour}, we can see how higher interferer power (lower $\rho$) changes the $\phi_k$ values at various locations in the network. 

From the above discussion, it is clear that if the central processing node has the knowledge of $\phi_k$ at each CR, it can set a threshold $\phi_0$ on the FVC value and eliminate CRs for which $\phi_k > \phi_0$, since they are closer to the interferer. Thus in the improved Cyclic WCL, each CR computes $\phi_k$ from the received signal according to (\ref{eq:phi_k}) and sends out this information to the central processing node. The CRs do not require the knowledge of $\rho$ and $\rho_k$ to compute $\phi_k$. 

It should be noted that in (\ref{eq:phi_k}), $\phi_k$ depends on ${\operatorname{var} ({{\hat R}_{{r_k}}^{\alpha_t}})}$ and ${E[|{{\hat R}_{{r_k}}^{\alpha_t}}{|^2}]}$. In a practical system, the $k^{th}$ CR estimates the variance of ${{\hat R}_{{r_k}}^{\alpha_t}}$ and mean of $|{{\hat R}_{{r_k}}^{\alpha_t}}{|^2}$ using $M$ realizations of ${\hat R}_{{r_k}}^{\alpha_t}$. The FVC at $k^{th}$ CR is an estimate based on $M$ realizations of ${\hat R}_{{r_k}}^{\alpha_t}$ and is denoted by $\hat{\phi}_k^M$. Therefore, we have
\begin{align}
{\hat{\phi}_k^M} &= \frac{v_s}{e_s},
\label{eq:sample_fvc}
\end{align}
where $v_s$ is sample variance given by ${v_s} = \frac{1}{M-1}\sum_{i=1}^{M}|(\hat{R}_{r_k}^{\alpha_t})_i-m_s|^2$ with $m_s = \frac{1}{M}\sum_{i=1}^{M}(\hat{R}_{r_k}^{\alpha_t})_i$ and $e_s  = \frac{1}{M}\sum_{i=1}^{M}|(\hat{R}_{r_k}^{\alpha_t})_i|^2= \frac{M-1}{M}v_s + |m_s|^2$. Here $(\hat{R}_{r_k}^{\alpha_t})_i$ is the $i^{th}$ realization of $\hat{R}_{r_k}^{\alpha_t}$. The number of realizations $M$ required to estimate accurate value of $\phi_k$ is obtained using confidence interval of the estimate $\hat{\phi}_k^M$. The analysis to find sufficient number of realizations $M$ for satisfactory performance of the algorithm is presented in Section \ref{sec:phi_k_M}. 

From the knowledge of $\hat{\phi}_k^M$, the central processing node estimates the target location by including CRs for which $\hat{\phi}_k^M\leq\phi_0$:
\begin{equation}
\mathbf{{\hat L}_{t_{improved}}}(\phi_0) = \frac{{\sum\limits_{k = 1}^K {|({\hat R_{{r_k}}^{\alpha_t}})_M{|^2}} \mathbf{L_k}} \mathds{1}(\hat{\phi}_k^M \leq \phi_0)}
{{\sum\limits_{k = 1}^K {|({\hat R_{{r_k}}^{\alpha_t}})_M{|^2}} \mathds{1}(\hat{\phi}_k^M \leq \phi_0)}},
\label{eq:lt_modified}
\end{equation}
where $({\hat R_{{r_k}}^{\alpha_t}})_M$ is the $M^{th}$ realization of ${\hat R_{{r_k}}^{\alpha_t}}$ and $\mathds{1}(\hat{\phi}_k^M \leq \phi_0)$ is an indicator function that is 1 for $\hat{\phi}_k^M\leq\phi_0$ and 0 otherwise. It should be noted that the locations of the CRs, the target and the interferer should remain constant until $MN$ samples of the received signal are collected at each CR. We assume that the locations remain constant for the time duration $MNT_s$. We also assume that received powers $p_{t,k}$ and $p_{i,k}$ remain constant for this duration.

In (\ref{eq:lt_modified}),  $\phi_0$ is a design parameter of the improved Cyclic WCL. Two important points should be noted while selecting the value of $\phi_0$. First, as observed in  Fig. \ref{fig:fvc_contour}, if $\phi_0$ is reduced for a fixed transmit power ratio, then CRs in a smaller area are included in Cyclic WCL. It means that only the CRs confined in the area where $\hat{\phi}_k^M \leq \phi_0$ are included in the Cyclic WCL. Second, if $\phi_0$ is fixed and the interferer power is increased, again CRs in a smaller area are included in the algorithm. Computation of the optimum value $\phi_0^{opt}$ of the FVC threshold that minimizes the RMSE is presented in Section \ref{sec:optimum_fvc_threshold}. This computation requires the information of $p_{t,k}$, $p_{i,k}$, $s_t$ and $s_i$. Since the central processing node might not have this information, it computes a suboptimal threshold $\phi_0^{sub}$ as described below.

Without the loss of generality, consider $\hat{\phi}_1^M \leq \hat{\phi}_2^M ...\leq \hat{\phi}_K^M $. It can be deduced from the analysis presented in Section \ref{sec:optimum_fvc_threshold}, that the unique values of $||\mathbf{\hat{L}_{t_{improved}}}(\phi_0)||^2$ occur at $\phi_0 \in \{\hat{\phi}_1^M, \hat{\phi}_2^M,...\hat{\phi}_K^M\}$, where $||.||$ is 2-norm of the vector.  Further, the optimal value of $\phi_0$ that minimizes the RMSE is one out of $\{\hat{\phi}_1^M, \hat{\phi}_2^M,...\hat{\phi}_K^M\}$. It should be noted that as the value of $\phi_0$ approaches $\phi_0^{opt}$, $||\mathbf{\hat{L}_{t_{improved}}}(\phi_0)||^2$ approaches $||\mathbf{{L}_{t}}||^2$. On the other hand, when the value of $\phi_0$ is significantly different from $\phi_0^{opt}$, $||\mathbf{\hat{L}_{t_{improved}}}(\phi_0)||^2$ takes values that are significantly different from $||\mathbf{{L}_{t}}||^2$. Based on this notion, the central processing node clusters $||\mathbf{\hat{L}_{t_{improved}}}(\phi_0)||^2, \phi_0 \in \{\hat{\phi}_1^M, \hat{\phi}_2^M,...\hat{\phi}_K^M\}$ in two sets: $\mathbb{S}^{opt}$ and $\mathbb{S}^{non-opt}$, using k-means algorithm \cite{Hartigan1979}. If $||\mathbf{\hat{L}_{t_{improved}}}(\phi_0)||^2 \in \mathbb{S}^{opt}$, then the value of $||\mathbf{\hat{L}_{t_{improved}}}(\phi_0)||^2$ is closer to  $||\mathbf{{L}_{t}}||^2$. Similarly if $||\mathbf{\hat{L}_{t_{improved}}}(\phi_0)||^2 \in \mathbb{S}^{non-opt}$, then  the value of $||\mathbf{\hat{L}_{t_{improved}}}(\phi_0)||^2$ is significantly different from $||\mathbf{{L}_{t}}||^2$. The set containing $||\mathbf{\hat{L}_{t_{improved}}}(\hat{\phi}_K^M)||^2$ is identified as $\mathbb{S}^{non-opt}$. This is due to the fact that for $\phi_0 = \hat{\phi}_K^M$, all the CRs, including those in the vicinity of the interferer are included in the localization which makes the value of $||\mathbf{\hat{L}_{t_{improved}}}(\phi_0)||^2$ differ significantly from $||\mathbf{{L}_{t}}||^2$. The other set obtained by k-means algorithm is then identified as $\mathbb{S}^{opt}$. Further the suboptimal FVC threshold $\phi_0^{sub}$ is computed as the average of  $\{ \phi_0 : ||\mathbf{\hat{L}_{t_{improved}}}(\phi_0)||^2 \in \mathbb{S}^{opt}\}$. 
The algorithm steps of the improved Cyclic WCL are described in Algorithm \ref{alg:improved_cyclic_wcl}.
\begin{algorithm}[t]
\caption{Improved Cyclic WCL}
\label{alg:improved_cyclic_wcl}
\begin{algorithmic}[1]
\State Estimate $\Delta\alpha = |\alpha_t - \alpha_i|$ and select $N > 10 \lceil \frac{f_s}{\Delta\alpha}\rceil$.
\For{i=1,2..M}
\State At each CR $k=1,2..K$, collect $N$ samples of the received signal $r_k(n), n=1,2..N$.
\State Compute CAC estimate: 
\Statex $({\hat R_{{r_k}}^{\alpha_t}})_i = \frac{1}
{N}\mathop \sum \limits_{n = 0}^{N-1} |{r_k}(n){|^2}{e^{ - j2\pi {\alpha _t}n{T_s}}}$
\EndFor
\State Using above $M$ realizations of ${\hat R_{{r_k}}^{\alpha_t}}$, compute $\hat{\phi}_k^M$ using (\ref{eq:sample_fvc}).
\State Compute the suboptimal FVC threshold $\phi_0^{sub}$ and estimate the target location using (\ref{eq:lt_modified}).
\end{algorithmic}
\end{algorithm}

\section{Theoretical Analysis}
\label{sec:Theoretical}
\subsection{Analysis of Cyclic WCL}
\label{sec:Theoretical_Cyclic_WCL}
In order to analyze the Cyclic WCL algorithm, we first formulate the estimates of the x- and y-coordinates of the target in the form of ratios of quadratic forms of a Gaussian vector (RQGV). The RMSE in the target location estimates is computed using the RQGV form.

Let us define Cyclic Cross-Correlation (CCC) between signals any two signals $u(n)$ and $v(n)$ at cyclic frequency $\alpha_t$ as
\begin{equation}
\begin{gathered}
  {{\hat R}_{uv}^{\alpha_t}} = \frac{1}
{N}\mathop \sum \limits_{n = 0}^{N-1} 2\operatorname{Re} \{ u(n)v{(n)^*}\} {e^{ - j2\pi {\alpha _t}n{T_s}}}. \hfill \\ 
\end{gathered} 
\label{eq:ccc_def}
\end{equation}
In the above equation, $(\hat{\text{       }})$ indicates the estimate based on $N$ samples and $(^*)$ indicates complex conjugate. Substituting (\ref{eq:rk}) in (\ref{eq:cac_rk}) and using the above definition, we can write the CAC of the received signal at $k^{th}$ CR as
\begin{align}
\nonumber
{{\hat R}_{{r_k}}^{\alpha_t}} ={\text{       }} &{p_{t,k}}{{\hat R}_{{s_t}}^{\alpha_t}} + {p_{i,k}}{{\hat R}_{{s_i}}^{\alpha_t}} + \sqrt {{p_{t,k}}{p_{i,k}}} {{\hat R}_{{s_t}{s_i}}^{\alpha_t}} \\&+ {{\hat R}_{{w_k}}^{\alpha_t}}  
  {\text{       }} + \sqrt {{p_{t,k}}} {{\hat R}_{{s_t}{w_k}}^{\alpha_t}} + \sqrt {{p_{i,k}}} {{\hat R}_{{s_i}{w_k}}^{\alpha_t}}.   
\label{eq:cac_rk_expanded}
\end{align}
Now, let us define three vectors: $\boldsymbol{\hat \theta_r} , \boldsymbol{\hat \theta_i}$ and $\mathbf{p_k}$. The vectors $\boldsymbol{\hat \theta_r}$ and $\boldsymbol{\hat \theta_i}$ contain real and imaginary parts of CAC and CCC, and $\mathbf{p_k}$ contains powers received at the $k^{th}$ CR from the target and the interferer. Therefore, we have
\begin{multline}
\nonumber
\boldsymbol{\hat \theta_r} = [\begin{matrix} {\operatorname{Re} \{ {{\hat R}_{{s_t}}^{\alpha_t}}\}, {\text{ }}\operatorname{Re} \{ {{\hat R}_{{s_i}}^{\alpha_t}}\}, {\text{ }}\operatorname{Re} \{ {{\hat R}_{{s_t}{s_i}}^{\alpha_t}}\}},\end{matrix}\\
\begin{matrix} \text{ }{\operatorname{Re} \{ {{\hat R}_{{w}}^{\alpha_t}}\} , 
{\text{ }}\operatorname{Re} \{ {{\hat R}_{{s_tw}}^{\alpha_t}}\} ,{\text{ }}\operatorname{Re} \{ {{\hat R}_{s_iw}^{\alpha_t}}\}}\end{matrix} ]^T,
\end{multline}
\begin{multline}
\nonumber
\boldsymbol{{\hat \theta_i }} = [\begin{matrix} {\operatorname{Im} \{ {{\hat R}_{{s_t}}^{\alpha_t}}\}, {\text{ }}\operatorname{Im} \{ {{\hat R}_{{s_i}}^{\alpha_t}}\}, {\text{ }}\operatorname{Im} \{ {{\hat R}_{{s_t}{s_i}}^{\alpha_t}}\}, }\end{matrix}\\
\begin{matrix}\text{ }{\operatorname{Im} \{ {{\hat R}_{{w}}^{\alpha_t}}\}, {\text{ }}\operatorname{Im} \{ {{\hat R}_{{s_tw}}^{\alpha_t}}\} ,{\text{ }}\operatorname{Im} \{ {{\hat R}_{s_iw}^{\alpha_t}}\}} \end{matrix}]^T,
\end{multline}
\begin{align}
\mathbf{p_k} &= {\left[ {{p_{t,k}},{\text{  }}{p_{i,k}},{\text{ }}\sqrt {{p_{t,k}}{p_{i,k}}}} ,\text{ } 1 ,\text{   }\sqrt {{p_{t,k}}},\text{  }\sqrt {{p_{i,k}}} \right]^T} .\hfill  
\label{eq:vector_def}
\end{align}
Using (\ref{eq:cac_rk_expanded}) and (\ref{eq:vector_def}),  we rewrite the estimate of x-coordinate of the target in terms of ratio of weighted sum of a vector norm:
\begin{equation}
{{\hat x}_t} = \frac{{\sum\limits_{k = 1}^K {|{{\hat R}_{{r_k}}^{\alpha_t}}{|^2}} {x_k}}}
{{\sum\limits_{k = 1}^K {|{{\hat R}_{{r_k}}^{\alpha_t}}{|^2}} }} = \frac{{{{\sum\limits_{k = 1}^K {\left\| {\left[ \begin{gathered}
  {\boldsymbol{\hat \theta_r }^T} \hfill \\
  {\boldsymbol{\hat \theta_i }^T} \hfill \\ 
\end{gathered}  \right]\mathbf{p_k}} \right\|} }^2}{x_k}}}
{{\sum\limits_{k = 1}^K {{{\left\| {\left[ \begin{gathered}
  {\boldsymbol{\hat \theta_r }^T} \hfill \\
  {\boldsymbol{\hat \theta_i }^T} \hfill \\ 
\end{gathered}  \right]\mathbf{p_k}} \right\|}^2}} }}.
\label{eq:xt}
\end{equation}
Further, we define power matrix $\mathbf{P} = [\mathbf{p_1, p_2,...p_K}]$ and position matrices $\mathbf{X} = \diag(x_1, x_2,...x_K)$ and $\mathbf{Y} = \diag(y_1, y_2,...y_K)$. From $\mathbf{P}$, $\mathbf{X}$ and $\mathbf{Y}$, symmetric matrices $\mathbf{A_x}$, $\mathbf{A_y}$, and $\mathbf{B}$ are obtained as shown below:
\begin{equation}
\begin{gathered}
  \mathbf{A_x} = \diag(\mathbf{PXP}^T, \mathbf{PXP}^T), \\
  \mathbf{A_y} = \diag(\mathbf{PYP}^T, \mathbf{PYP}^T),\\
  \mathbf{B} = \diag(\mathbf{PP}^T, \mathbf{PP}^T). 
\end{gathered}
\label{eq:matrix_def}
\end{equation}
As shown in the Appendix A, $\hat x_t$ can be written in terms of $\mathbf{A_x}$, $\mathbf{B}$ and $\boldsymbol{\hat \theta}  = {\left[ {{\boldsymbol{\hat \theta_r }^T}{\text{ }}{\boldsymbol{\hat \theta_i }^T}} \right]^T}$ as:
\begin{equation}
{{\hat x}_t} = \frac{{{\boldsymbol{\hat \theta }^T}\mathbf{A_x}\boldsymbol{\hat \theta} }}
{{{\boldsymbol{\hat \theta }^T}\mathbf{B}\boldsymbol{\hat \theta} }}.
\label{eq:xt_2}
\end{equation}
Similarly, the estimate of the y-coordinate can be written as:
\begin{equation}
{{\hat y}_t} = \frac{{{\boldsymbol{\hat \theta }^T}\mathbf{A_y}\boldsymbol{\hat \theta} }}
{{{\boldsymbol{\hat \theta }^T}\mathbf{B}\boldsymbol{\hat \theta} }},
\label{eq:yt_2}
\end{equation}

The target location estimates in (\ref{eq:xt_2}) and (\ref{eq:yt_2}) are functions of the vector $\boldsymbol{\hat \theta} = {\left[ {{\boldsymbol{\hat \theta_r }^T}{\text{ }}{\boldsymbol{\hat \theta_i }^T}} \right]^T}$ that contains real and imaginary parts of estimates of CACs and CCCs as defined in (\ref{eq:vector_def}). The estimates of CACs and CCCs are computed using $N$ samples of corresponding signals and are modeled as Gaussian random variables for a sufficiently large value of $N$ \cite[Eqn. (20)]{Dandawate1994}.
It follows that the vector $\boldsymbol{\hat \theta}$ is a Gaussian vector and is a function of number of samples observed $N$. 

As shown in Appendix B, CACs and CCCs are computed from moments of $s_t$ and $s_i$, which in turn are functions of data symbols, $a_l, b_l$ and pulse shapes $g_{n,l}, h_{n,l}$ for single carrier signals. For OFDM signals, the moments  are functions of subcarrier symbols $c_{\kappa,l}, d_{\kappa,l}$, subcarrier spacings $\Delta f_t, \Delta f_i$, and window functions $g_{n,l}, h_{n,l}$. Therefore, the mean $E[\boldsymbol{\hat{\theta}}]$ and the covariance matrix $\boldsymbol{\Sigma_{\hat \theta}}$ of the Gaussian vector $\boldsymbol{\hat \theta}$ are derived in terms of these parameters as shown in Appendix B.
The target location estimates in (\ref{eq:xt_2}) and (\ref{eq:yt_2}) also depend on the locations of CRs and the power received at the CRs through matrices $\mathbf{A_x}$, $\mathbf{A_y}$, and $\mathbf{B}$.

In order to compute the RMSE, we find the second moments of location estimates $E[\hat{x}_t^2]$ and $E[\hat{y}_t^2]$. It should be noted that $\hat{x}_t$ and $\hat{y}_t$ are in the RQGV form with vector $\boldsymbol{\hat \theta}$. The second moments of RQGV are given in  \cite[Thm. 6]{Magnus1986}. To utilize the result presented in \cite{Magnus1986}, the matrix $\mathbf{B}$ should be positive semidefinite. By definition, $\mathbf{B} = \diag(\mathbf{P}\mathbf{P^T}, \mathbf{P}\mathbf{P^T})$. Any matrix of the form $\mathbf{P}\mathbf{P^T}$ is positive semi-definite. Since $\mathbf{B}$ has  $\mathbf{P}\mathbf{P^T}$ on its diagonal, $\mathbf{B}$ is also a positive  semidefinite matrix. 

First, we compute Cholesky factorization of $\boldsymbol{\Sigma_{\hat \theta}}= \mathbf{CC^T}$. Then, the eigenvalue decomposition of $\mathbf{C^TBC}$ gives the orthogonal matrix $\mathbf{V}$ with eigenvectors on its columns and diagonal matrix $\boldsymbol{\Lambda}$ with eigenvalues on the diagonal. Define a matrix $\mathbf{A^*}=\mathbf{V^TC^TA_xCV}$ and a vector $\boldsymbol{\mu}=\mathbf{V^TC^{-1}E[\boldsymbol{\hat{\theta}}]}$. Further,  $E[\hat{x}_t^2]$ is computed, using \cite[Thm. 6]{Magnus1986}, as:
\begin{align}
\nonumber
E[{{\hat x}^2}] = E\left[ {{{\left( {\frac{{{\boldsymbol{\hat \theta }^T}\mathbf{A_x}\boldsymbol{\hat \theta} }}
{{{\boldsymbol{\hat \theta }^T}\mathbf{B}\boldsymbol{\hat \theta }}}} \right)}^2}} \right]= {e^{ - \frac{1}
{2}E{{[\boldsymbol{\hat \theta} ]}^T}\boldsymbol{\Sigma _{\hat \theta }}^{ - 1}E[\boldsymbol{\hat \theta} ]}}\times \\\nonumber
\left\{ {2\int\limits_0^\infty  {t \text{ } \operatorname{det}(\boldsymbol{\Delta} ){e^{\frac{1}
{2}\boldsymbol{\zeta ^T}\zeta }}} \left( {\operatorname{Tr}(\mathbf{R^2}) + 2\boldsymbol{\zeta ^T}\mathbf{R^2}\boldsymbol{\zeta} } \right)dt} \right. \hfill \\
  \left. { + \int\limits_0^\infty  {t \text{ } \operatorname{det}(\boldsymbol{\Delta} ){e^{\frac{1}
{2}\boldsymbol{\zeta ^T}\boldsymbol{\zeta} }}} {{\left( {\operatorname{Tr}(\mathbf{R}) + \boldsymbol{\zeta ^T}\mathbf{R}\boldsymbol{\zeta} } \right)}^2}dt} \right\} \hfill
\label{eq:e_xt_2}
\end{align}
where, $\boldsymbol{\Delta}=(\mathbf{I_n}+2t\boldsymbol{\Lambda})^{-\frac{1}{2}}$, $\mathbf{R}=\mathbf{\Delta A^* \Delta}$, $\boldsymbol{\zeta} = \boldsymbol{\Delta\mu}$.

Similarly, $E[\hat{y}_t^2]$ is obtained by replacing $\mathbf{A_x}$ with $\mathbf{A_y}$. From the second moments, we get the theoretical value of the RMSE ($\epsilon$) using
\begin{equation}
\epsilon= \sqrt{E[\hat{x}_t^2]+E[\hat{y}_t^2]}.
\label{eq:rmse}
\end{equation}
\subsection{Analysis of Improved Cyclic WCL}
\label{sec:Theoretical_Improved_Cyclic_WCL}
\subsubsection{Estimation of Feature Variation Coefficient $\phi_k$}
\label{sec:phi_k_M}
As mentioned in Section \ref{sec:Improved Cyclic WCL}, in the practical system, the $k^{th}$ CR estimates the value of $\phi_k$ using $M$ realizations. We denote the estimate of $\phi_k$ using $M$ realizations by $\hat{\phi}_k^M$ and it is given by $\hat{\phi}_k^M = v_s/e_s$, where $v_s$ and $e_s$ are sample variance of $\hat{R}_{r_k}^{\alpha_t}$ and sample mean of $|\hat{R}_{r_k}^{\alpha_t}|^2$, respectively.

The sample variance ${v_s}$ follows the Chi-square distribution, but it can be approximated by the Gaussian distribution for $M>50$ \cite[pp.118]{Box1978}. We consider $M>50$ for simplicity of analysis. The mean ($\mu_{v_{s}}$) and variance ($\sigma_{v_{s}}^2$) of $v_s$ are given by \cite{Garcia2008}
\begin{align}
\mu_{v_{s}} = \operatorname{var}(\hat{R}_{r_k}^{\alpha_t}) \text{\ and }
\sigma_{v_{s}}^2 = \frac{1}{M}\left[{\mu_4 - \frac{M-3}{M-1}\mu_{v_{s}}^2}\right],
\label{eq:mean_var_v_s}
\end{align}
where $\mu_4 = E\{|\hat{R}_{r_k}^{\alpha_t} - E[\hat{R}_{r_k}^{\alpha_t}]|^4\}$.
Similarly, ${e_s}$ is a Gaussian random variable with mean $\mu_{e_{s}}$ and variance $\sigma_{e_{s}}^2$ as follows:
\begin{align}
\mu_{e_{s}} = E[|\hat{R}_{r_k}^{\alpha_t}|^2]  , \text{\ and }
\sigma_{e_{s}}^2=\operatorname{var}(|\hat{R}_{r_k}^{\alpha_t}|^2)/M.
\label{eq:mean_var_e_s}
\end{align}
The analytical expressions of $\mu_{v_s}$, $\sigma_{v_s}$, $\mu_{e_s}$, $\sigma_{e_s}$ in terms of $E[\boldsymbol{\hat{\theta}}]$, $\boldsymbol{\Sigma_{\hat{\theta}}}$ and $\mathbf{p_k}$ are presented in Appendix E. It should be noted that $\sigma_{v_s}^2$ and $\sigma_{e_s}^2 \rightarrow 0$  as $M \rightarrow \infty$ since $v_s$ and $e_s$ are consistent estimators of $\operatorname{var}(\hat{R}_{r_k}^{\alpha_t})$ and $E[|\hat{R}_{r_k}^{\alpha_t}|^2]$.


We compute the confidence interval for $\hat{\phi}_k^M$ using the fact that it is a ratio of two Gaussian variables $v_s$ and $e_s$. Let $\beta$ be the required confidence level, i.e., the probability that the true value of $\phi_k$ lies within the given confidence interval. Further, let $C$ be the center of the confidence interval corresponding to the confidence level $\beta$ and $S$ be the standard error in the computation of $\phi_k$. Then from \cite{Dunlap1986}, we get
\begin{gather}
\nonumber
C = \frac{\frac{v_s}{e_s} - z_\beta^2 \frac{\sigma_{v_se_s}}{e_s^2}}{Q} \text{,}  \\
S^2 = \frac{\sigma_{v_s}^2 - 2 \frac{v_s}{e_s}\sigma_{v_se_s}+\frac{v_s^2}{e_e^2}\sigma_{e_s}^2 - z_\beta^2 \frac{\sigma_{e_s}^2}{e_s^2}\left[\sigma_{v_s}^2 - \frac{\sigma_{v_se_s}^2}{\sigma_{e_s}^2}\right]}{e_s^2Q^2}
\label{eq:confidence_interval_1}
\end{gather}
and the confidence interval is $C.I. = C \pm z_\beta S$, where $Q = 1 - z_\beta^2 \sigma_{e_s}^2 / e_s^2$ and $z_\beta$ is Student's-t variable corresponding to the confidence level $\beta$ and the number of realizations $M$. The value of $z_\beta$ is obtained from standard tables such as \cite[Table 8.2]{Garcia2008}. 

It should be noted that $z_\beta S$ reduces with if the $M$ is increased. While computing $\hat{\phi}_k^M$, the number of realizations $M$ should be large enough to satisfy $z_\beta S < \delta$ for a small value of $\delta$. The value of $\delta$ is selected as the minimum difference between the FVC at any two CRs, i.e., $\delta = \min \limits_{i,j}({\phi_i}- {\phi_j}),$ for $i,j \in \{1,2,...K\}, i \neq j$. This value of $\delta$ ensures that only CRs for which $\phi_k \leq \phi_0$ are included in the algorithm with confidence level $\beta$. In other words, if $\phi_k \leq \phi_0$, then we have $\hat{\phi}_k^M\leq \phi_0$. Other other hand, if $\phi_k > \phi_0$, then we have $\hat{\phi}_k^M> \phi_0$ with probability $\beta$.
 


\subsubsection{RMSE in the Improved Cyclic WCL Estimates}
\label{sec:Rmse_Improved_Cyclic_WCL}
As mentioned in the previous section, the improved Cyclic WCL includes only CRs with $\hat{\phi}_k^M \leq \phi_0$, where $\phi_0$ is the FVC threshold. In order to write the estimates of x- and y-coordinates as a function of $\phi_0$, we introduce a threshold based $K \times K$ diagonal matrix, called selection matrix $\mathbf{S_0}$. The $k^{th}$ diagonal element of $\mathbf{S_0}$ is 1 if $\hat{\phi}_k^M \leq \phi_0$ and 0 otherwise. In other words, the $k^{th}$ diagonal element of the matrix $\mathbf{S_0}$ is the indicator function $\mathds{1}(\hat{\phi}_k^M\leq\phi_0)$. The selection matrix is incorporated in previous definitions of symmetric matrices in (\ref{eq:matrix_def}) as 
\begin{align}
\nonumber  \mathbf{A_x'} &= \diag(\mathbf{PS_0XS_0}^T\mathbf{P}^T,\mathbf{PS_0XS_0}^T\mathbf{P}^T) \hfill, \\
\nonumber  \mathbf{A_y'} &= \diag(\mathbf{PS_0YS_0}^T\mathbf{P}^T,\mathbf{PS_0YS_0}^T\mathbf{P}^T) \hfill \text{, and} \\
  \mathbf{B'} &= \diag(\mathbf{PS_0S_0}^T\mathbf{P}^T, \mathbf{PS_0S_0}^T\mathbf{P}^T). \hfill 
\label{eq:matrix_def_improved}
\end{align}
The location estimates using the improved Cyclic WCL are given as
\begin{equation}
{{\hat x}_{t}(\phi_0)} = \frac{{{\boldsymbol{\hat \theta }^T}\mathbf{A_x'}\boldsymbol{\hat \theta} }}
{{{\boldsymbol{\hat \theta }^T}\mathbf{B'}\boldsymbol{\hat \theta} }}{\text{ and }}{{\hat y}_{t}(\phi_0)} = \frac{{{\boldsymbol{\hat \theta }^T}\mathbf{A_y'}\boldsymbol{\hat \theta} }}
{{{\boldsymbol{\hat \theta }^T}\mathbf{B'}\boldsymbol{\hat \theta} }}.
\label{eq:xt_yt_improved}
\end{equation}

It should be noted that ${{\hat x}_{t}(\phi_0)}$ and ${{\hat y}_{t}(\phi_0)}$ are also in RQGV form in the Gaussian vector $\boldsymbol{\hat{\theta}}$. Further, in order to show that $\mathbf{B'}$ is positive semidefinite, we note that $\mathbf{PS_0S_0}^T\mathbf{P}^T$ is positive semi-definite, if there is at least one non-zero element on the diagonal of $\mathbf{S_0}$. The number of non-zero elements of the diagonal of $\mathbf{S_0}$ is the number of CRs satisfying $\hat{\phi}_k^M \leq \phi_0$. If $\phi_0$ is selected such that at least one CR satisfies $\hat{\phi}_k^M \leq \phi_0$, then $\mathbf{PS_0S_0}^T\mathbf{P}^T$ is positive semi-definite, which in turn makes $\mathbf{B'}$ also a positive semi-definite matrix. Therefore, for a fixed value of $\phi_0$, the RMSE for improved Cyclic WCL is computed in similar way as Cyclic WCL: 
\begin{equation}
\epsilon(\phi_0) = E[\hat{x}_{t}^2 (\phi_0)] + E[\hat{y}_{t}^2 (\phi_0)],
\label{eq:rmse_improved}
\end{equation}
where  $E[\hat{x}_t^2 (\phi_0)]$  and $E[\hat{y}_t^2 ( \phi_0)]$ are obtained by replacing $\mathbf{A_x}$, $\mathbf{A_y}$ and $\mathbf{B}$ with $\mathbf{A_x'}$ and $\mathbf{A_y'}$ and $\mathbf{B'}$, respectively in (\ref{eq:e_xt_2}).

\subsubsection{Optimum FVC threshold}
\label{sec:optimum_fvc_threshold}
The optimum value of the FVC threshold $\phi_0$ that minimizes the RMSE $\epsilon(\phi_0)$ is obtained with the knowledge of $\mathbf{A_x'}$, $\mathbf{A_y'}$, $\mathbf{B'}$, $E[\boldsymbol{\hat{\theta}}]$ and $\boldsymbol{\Sigma_{\hat \theta}}$. Without loss of generality, let us consider  $\hat{\phi}_1^M \leq \hat{\phi}_2^M ...\leq \hat{\phi}_K^M$. For $\phi_0 < \hat{\phi}_1^M$, no CR will be used for localization and the target location estimates cannot be computed. Therefore, we must have $\phi_0 \geq \hat{\phi}_1^M$ for the improved Cyclic WCL algorithm to work. If the value of FVC threshold is $\phi_0 = \hat{\phi}_{k'}^M$ $(1\leq k' \leq K)$, then CRs $1,2,3..k'$  will be included in the localization process.

It should be noted that the RMSE for any value of $\phi_0$ in the range $[\hat{\phi}_{k'}^M, \hat{\phi}_{k'+1}^M)$ remains constant. This is because the same $k'$ CRs are used for localization if the value of $\phi_0$ is in the given range. Therefore, $\epsilon(\phi_0) = \epsilon(\hat{\phi}_{k'}^M) $ if $\phi_0 \in [\hat{\phi}_{k'}^M,\hat{\phi}_{k'+1}^M)$. It follows from the above argument that $\epsilon({\phi_0})$ has $K$ unique values in the domain of $\phi_0 \in [0,1]$ and the unique values are  $\{\epsilon(\hat{\phi}_{1}^M), \epsilon(\hat{\phi}_{2}^M),....\epsilon(\hat{\phi}_{K}^M)\}$. We obtain the unique values of $\epsilon({\phi_0})$ using (\ref{eq:rmse_improved}) at $\phi_0 \in \{\hat{\phi}_{1}^M, \hat{\phi}_{2}^M,...\hat{\phi}_{K}^M\}$. The optimum value of the FVC threshold is then given by $\{\phi_0^{opt}: \epsilon(\phi_0^{opt}) \leq \epsilon(\hat{\phi}_k^M), k=1,2..K \}$.


\subsection{{Complexity Analysis}}
\label{sec:complexity}
In this section, we compare the computational complexities of the proposed algorithms, Cyclic WCL and improved Cyclic WCL, with traditional WCL algorithm. We assume that the number of operations (OPS) required for addition, subtraction and comparison is 1, and for multiplication and division is 10, as considered in \cite{Wang2011}. The number of OPS required for Cyclic WCL, using (\ref{eq:rk}) and (\ref{eq:lt}), is $21KN + 23K + 36$. Similarly, the number of OPS required for improved Cyclic WCL, using (\ref{eq:rk}), (\ref{eq:phi_k}), and (\ref{eq:lt_modified}), is $21MKN+24MK+19M+(71+\eta) K+17$. Here, $\eta K$ is the number of OPS required to obtain $\phi_0^{sub}$ using k-means clustering. The computational complexity of the k-means algorithm is $\mathcal{O}(K)$ \cite{Hamerly2015}. Therefore, we assume $\eta K$ is the number of OPS required for k-means clustering. 

Finally, the WCL algorithm in \cite{Wang2011} is a special case of Cyclic WCL with $\alpha_t=0$. The number of OPS required for WCL is $11KN+23K+26$. Therefore, the computational complexities of WCL and Cyclic WCL are $\mathcal{O}(KN)$, while that of improved Cyclic WCL is $\mathcal{O}(MKN)$.

\begin{figure}[t!]
	\centering
	\includegraphics[width=0.9\columnwidth]{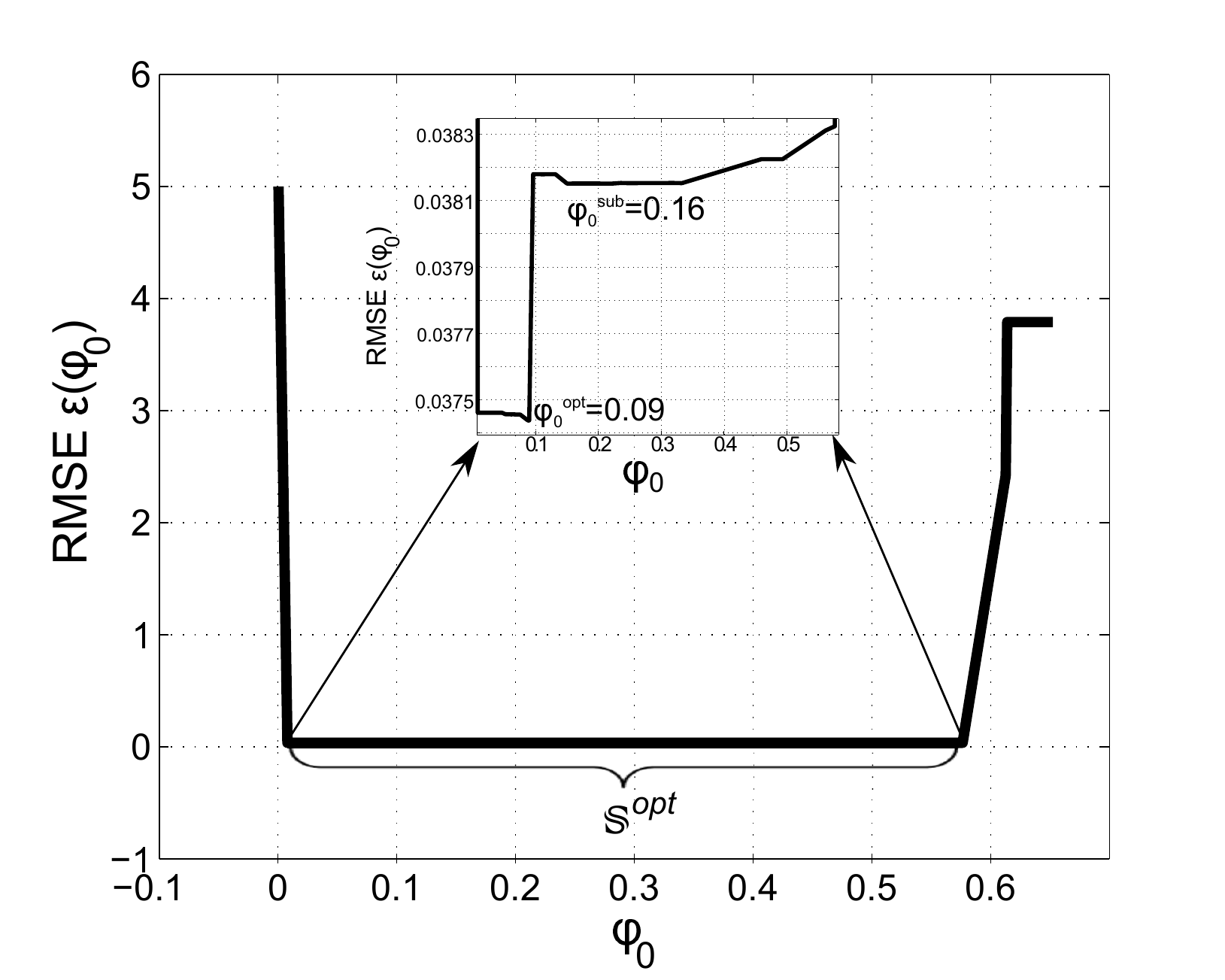}
	\caption{Improved Cyclic WCL: RMSE vs $\phi_0$. $\phi_0^{opt}=0.09, \epsilon(\phi_0^{opt})=0.03744 $. $\phi_0^{sub}=0.16, \epsilon(\phi_0^{sub})=0.03815$. $K = 50$. $\rho = -10$ dB. Locations of the CRs [$x_k, y_k$] and $x_k \in \{-40, -20, 0, 20, 40\}$ and $y_k \in \{-45, -35, -25, -15, -5, 5 , 15, 25, 35, 45\}$. Interferer location: $[20, 20]$.}
	\label{fig:rmse_vs_phi_0}
\end{figure}

\section{Simulation Results and Discussion}
\label{sec:Results}
We consider a CR network with $K$ CRs located in a square shaped area of size $100$m x $100$m. The path-loss exponent $\gamma = 3.8$ and noise PSD is $N_0 = -174$dBm/Hz. Initially, we show results with single carrier 4-QAM signals with carrier frequency $f_t=f_i = 2.4$GHz. The symbol rates of the target and  the interferer signals are $\alpha_t=20$MHz and $\alpha_i=25$MHz, respectively. The sampling frequency is $f_s = 200$MHz. The number of samples $N$ is selected such that $N>10 \lceil \frac{f_s}{\Delta\alpha}\rceil = 400$.

First, improved Cyclic WCL is studied in Section \ref{sec:Results_Improved_Cyclic_WCL}. Then, we compare the performances of improved Cyclic WCL and Cyclic WCL in Section \ref{sec:Results_Comparison_Cyclic_WCLs} in the absence of shadowing ($\sigma_q = 0$dB). The comparison between the traditional WCL, Cyclic WCL and the improved Cyclic WCL under the shadowing environment is presented in Section \ref{sec:Results_Comparison}. This section also shows results with OFDM signals. Finally, simulation results with multipath fading channels are shown in Section \ref{sec:Results_multipath}.
\subsection{Performance of the improved Cyclic WCL}
\label{sec:Results_Improved_Cyclic_WCL}
In  improved Cyclic WCL, the number of realizations $M$ are computed using confidence interval as described in Section \ref{sec:phi_k_M}. Further, following the analysis in Section \ref{sec:Rmse_Improved_Cyclic_WCL}, the RMSE in the localization estimates is obtained. Note that above analysis holds for given fixed locations of the CRs, the target and the interferer. Therefore, in order to evaluate the performance of the improved Cyclic WCL, first we show the simulation results for fixed locations of the CRs in Section \ref{sec:Results_Fixed_Locations}. In Section \ref{sec:Results_Uniform_Locations}, we consider uniformly distributed CRs in the network and compute the average RMSE over 1000 iterations.

\begin{figure}[t!]
	\centering
	\includegraphics[width=0.9\columnwidth]{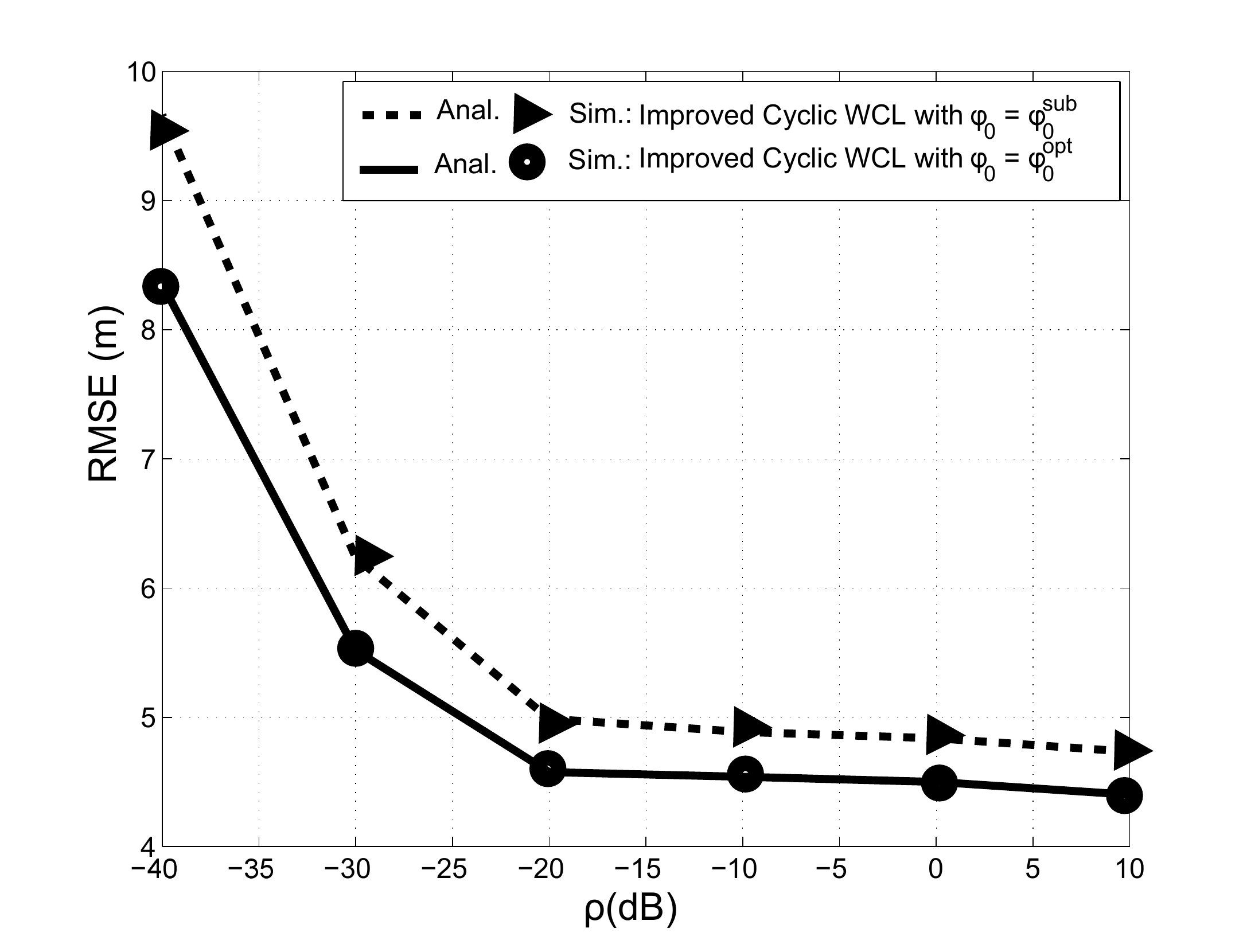}
	\caption{RMSE vs transmit power ratio ($\rho$) for improved Cyclic WCL with optimal and suboptimal FVC threshold. Number of CRs $K=50$. Number of Samples $N=500$. Interferer location: $[x_i,y_i]= [20, 20]$.}
	\label{fig:rmse_vs_rho_opt_subopt}
\end{figure}

\subsubsection{Performance of improved Cyclic WCL for fixed locations of CRs}
\label{sec:Results_Fixed_Locations}
The locations of $K=50$ CRs are fixed at $[x_k, y_k]$ such that $x_k \in \{-40, -20, 0, 20, 40\}$ and $y_k \in \{-45, -35, -25, -15, -5, 5 , 15, 25, 35, 45\}$. The target and the interferer are located at $[0, 0]$ and $[20, 20]$, respectively.

The number of realizations $M$ required to obtain $\hat{\phi}_k^M$ are computed with confidence level $\beta = 0.9$ and $\delta = 0.01$. 
For the given parameters, it was observed that $z_\beta S < 0.01$ for $M \geq 60$, therefore $M=60$ realizations are used to compute $\hat{\phi}_k^M$. Further, the optimum threshold $\phi_0$ was obtained as discussed in Section \ref{sec:optimum_fvc_threshold}. In the scenario considered here, $\phi_0^{opt} = 0.09$ and the corresponding RMSE is $\epsilon(\phi_0^{opt})=0.03744$. The impact of selecting different FVC thresholds on the error is shown in Fig. \ref{fig:rmse_vs_phi_0}. The y-axis in this figure also represents $||\boldsymbol{\hat{L}_{t_{improved}}}(\phi_0)||^2$ since the target is located at the origin. In this particular case, the k-means clustering results in $||\boldsymbol{\hat{L}_{t_{improved}}}(\phi_0)||^2 \in \mathbb{S}^{opt}$ for $0.08 \leq \phi_0 \leq 0.57$. Further, the suboptimal threshold $\phi_0^{sub} = 0.16$ and the corresponding RMSE is $\epsilon(\phi_0^{sub})=0.03815$. Hence, the localization error increased by only 0.0007m if suboptimal threshold is used in this setting.
\subsubsection{Performance of improved Cyclic WCL with uniformly distributed CRs}
\label{sec:Results_Uniform_Locations}
Now we consider that the CRs are uniformly distributed in the network. The average RMSE of 1000 realizations of CR locations is plotted against the transmit power ratio ($\rho$) in Fig. \ref{fig:rmse_vs_rho_opt_subopt}. The figure compares performance of the algorithm with optimal ($\phi_0^{opt}$) and suboptimal ($\phi_0^{sub}$) FVC threshold. As described in Section \ref{sec:Improved Cyclic WCL}, unlike $\phi_0^{opt}$,  the computation of $\phi_0^{sub}$ does not require the knowledge of $p_{t,k}, p_{i,k}, s_t$ and $s_i$. From Fig. \ref{fig:rmse_vs_rho_opt_subopt}, it can be observed that the performance of the improved Cyclic WCL with suboptimal FVC threshold is comparable to the performance with optimal threshold. The suboptimal threshold results in increased error of at most $1$m for transmit power ratio ranging from $10$ dB to $-40$dB. Therefore, the knowledge of the transmit powers of the target and the interferer and their signals is not necessary for satisfactory performance of the proposed coarse-grained localization algorithm.
\subsection{Comparison between Cyclic WCL and improved Cyclic WCL}
\label{sec:Results_Comparison_Cyclic_WCLs}
In this section, the impacts of the location of the interferer, CR density and the number of samples $N$ on Cyclic WCL and improved Cyclic WCL are studied. In each scenario, the algorithm is simulated for 1000 realizations with uniformly distributed CRs. The average RMSE over 1000 realizations is plotted. The improved Cyclic WCL is implemented with suboptimal threshold $\hat{\phi}_0^{sub}$.

\begin{figure}[t!]
	\centering
	\begin{subfigure}[c]{0.45\textwidth}
		\includegraphics[width=0.9\columnwidth]{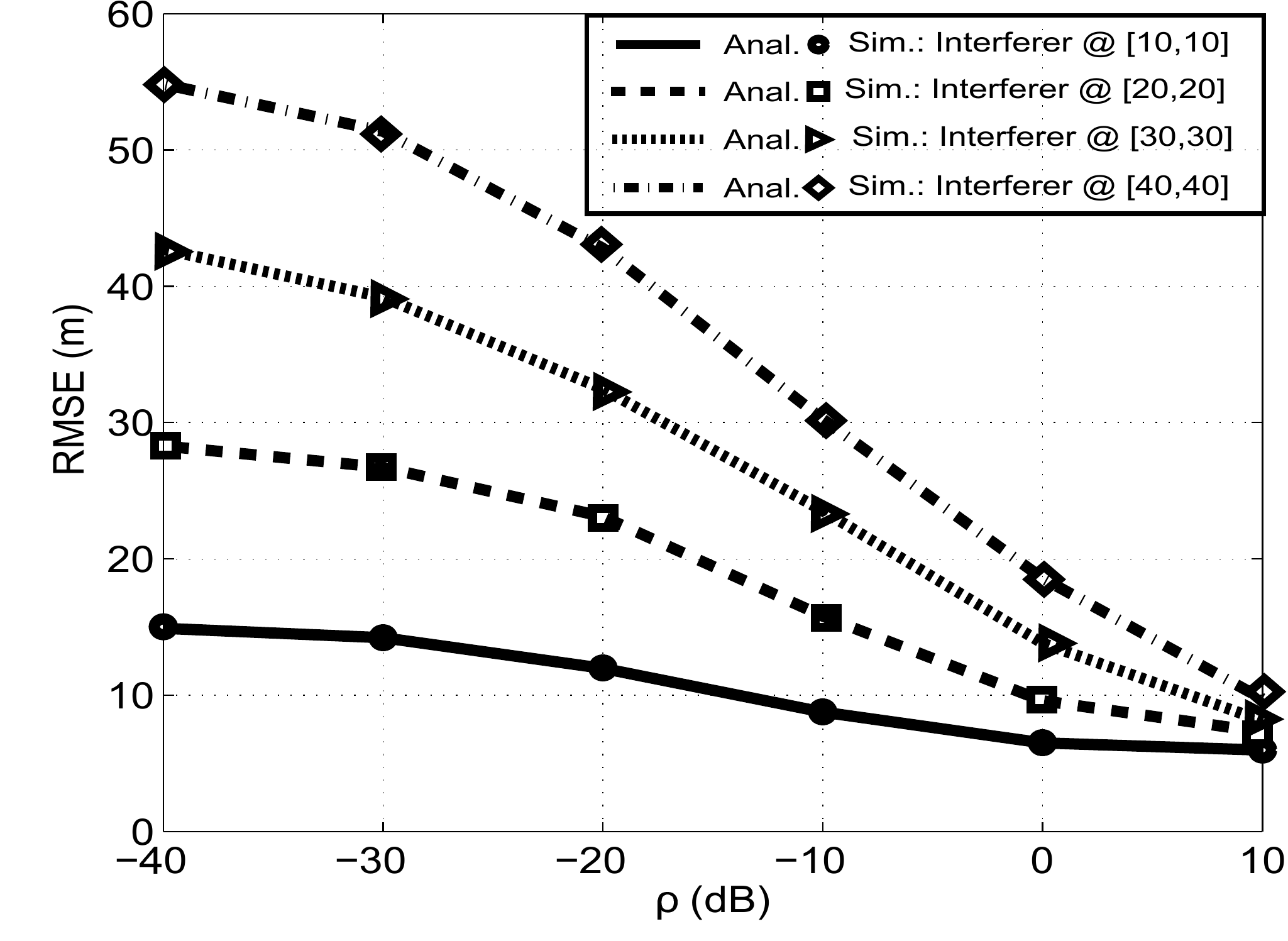}
		\caption{Cyclic WCL}
		\label{fig:diff_loc_cwcl}
	\end{subfigure}	
	\begin{subfigure}[c]{0.5\textwidth}
		\includegraphics[width=0.9\columnwidth]{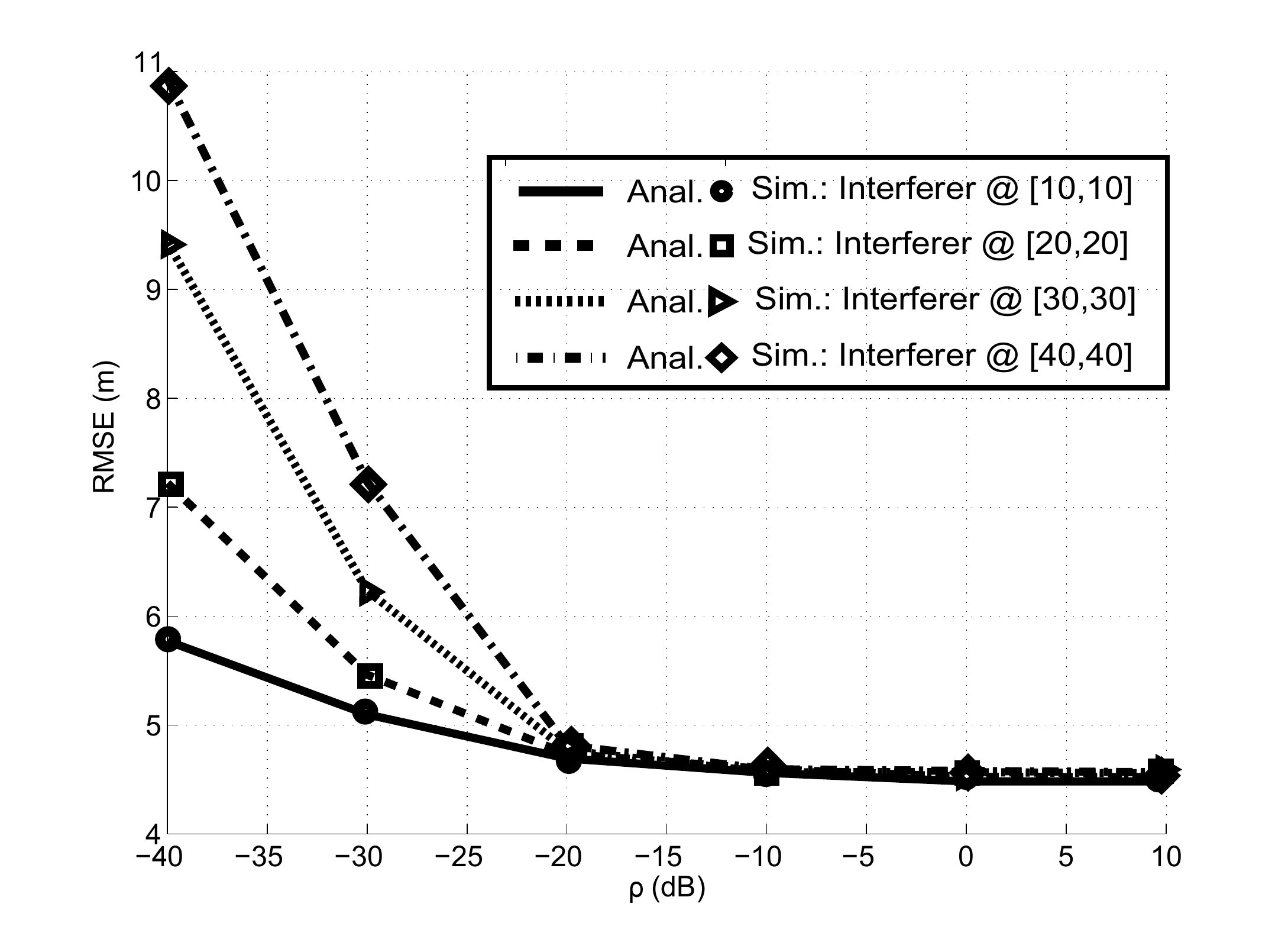}
		\caption{{Improved Cyclic WCL}}
		\label{fig:diff_loc_icwcl}
	\end{subfigure}
	\caption{Impact of interferer location on a) Cyclic WCL and b) Improved Cyclic WCL, with $p_t=10$dBm. Number of CRs $K= 50$. Number of samples $N=500$. Target location: $[0, 0]$.}
	\label{fig:rmse_vs_rho_diff_loc}
\end{figure}

\subsubsection{Impact of the location of the interferer}
The RMSE in the target location estimates for different locations of the interferer is shown in Fig. \ref{fig:rmse_vs_rho_diff_loc}. The results shown in the figures are counter-intuitive, since the interferer located further away from the target causes higher error as compared to the interferer located closer to the target, especially with increased interferer power. This is due to the fact that at high interferer power, the centroid of $|\hat{R}_{r_k}|^2$ is closer to the interferer. Further, if the interferer location is away from the target, the centroid and hence the target location estimates move away from the target and closer to the interferer. This phenomenon results in increased error as seen in the figure. For a fixed position of the interferer, it is observed that the RMSE increases with higher interferer power, since the impact of the interferer on the centroid of $|\hat{R}_{r_k}|^2$ becomes more prominent. {As shown in Fig.} \ref{fig:diff_loc_cwcl} {and} \ref{fig:diff_loc_icwcl}, {the RMSE in improved Cyclic WCL is significantly lower than RMSE in Cyclic WCL. Further, the RMSE for different locations converge at $\rho=10$ dB for Cyclic WCL, and at $\rho= -20$dB for improved Cyclic WCL. Therefore, improved Cyclic WCL provides more robustness against interferer's power and location.}

\begin{figure} [t!]
	\centering
	\begin{subfigure} [b]{0.45\textwidth}
		\includegraphics[width=\textwidth]{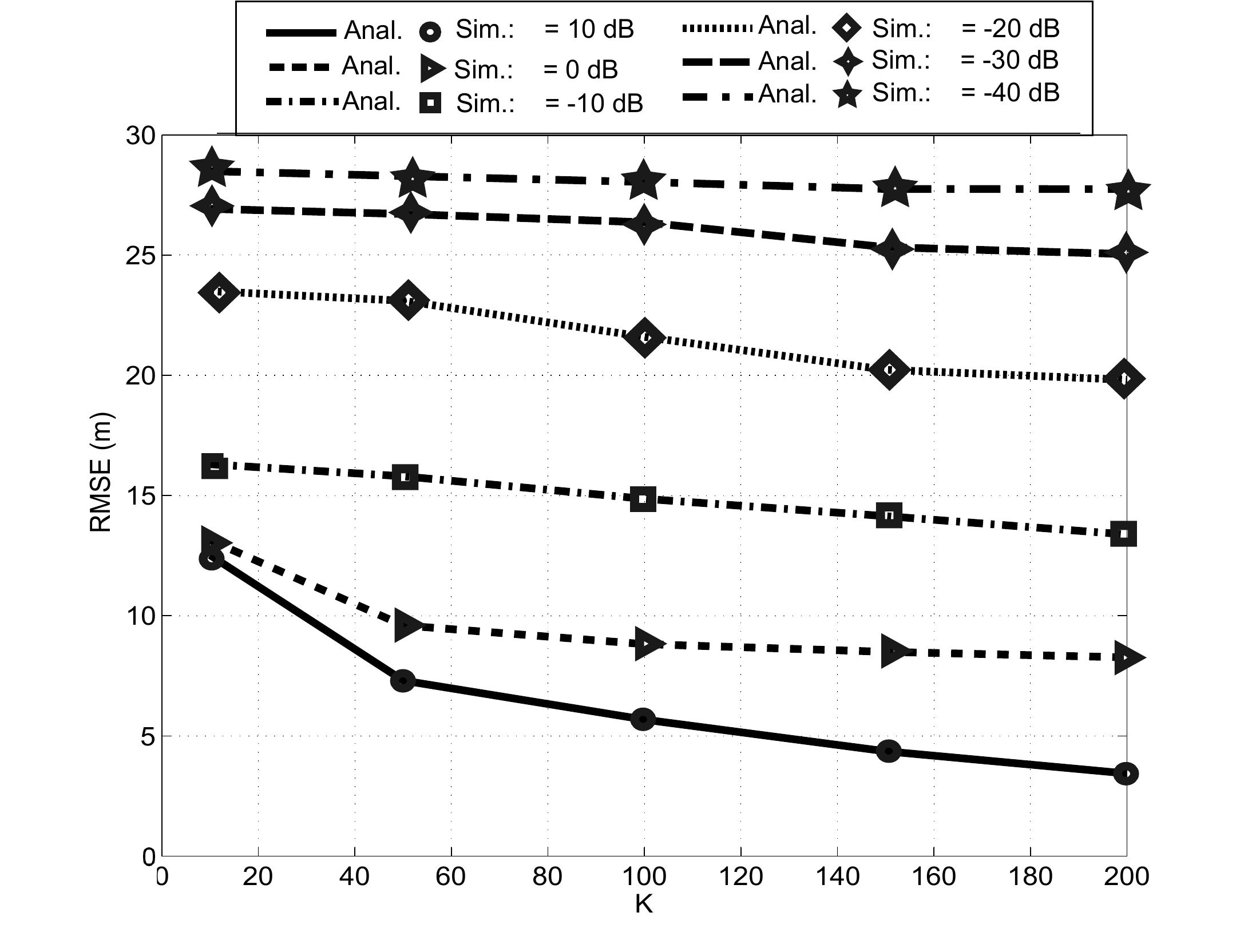}
		\caption{Cyclic WCL}
		\label{fig:diff_K_cwcl}
	\end{subfigure}
	\hspace{0.5em}
	\begin{subfigure}[b]{0.45\textwidth}
		\includegraphics[width=\textwidth]{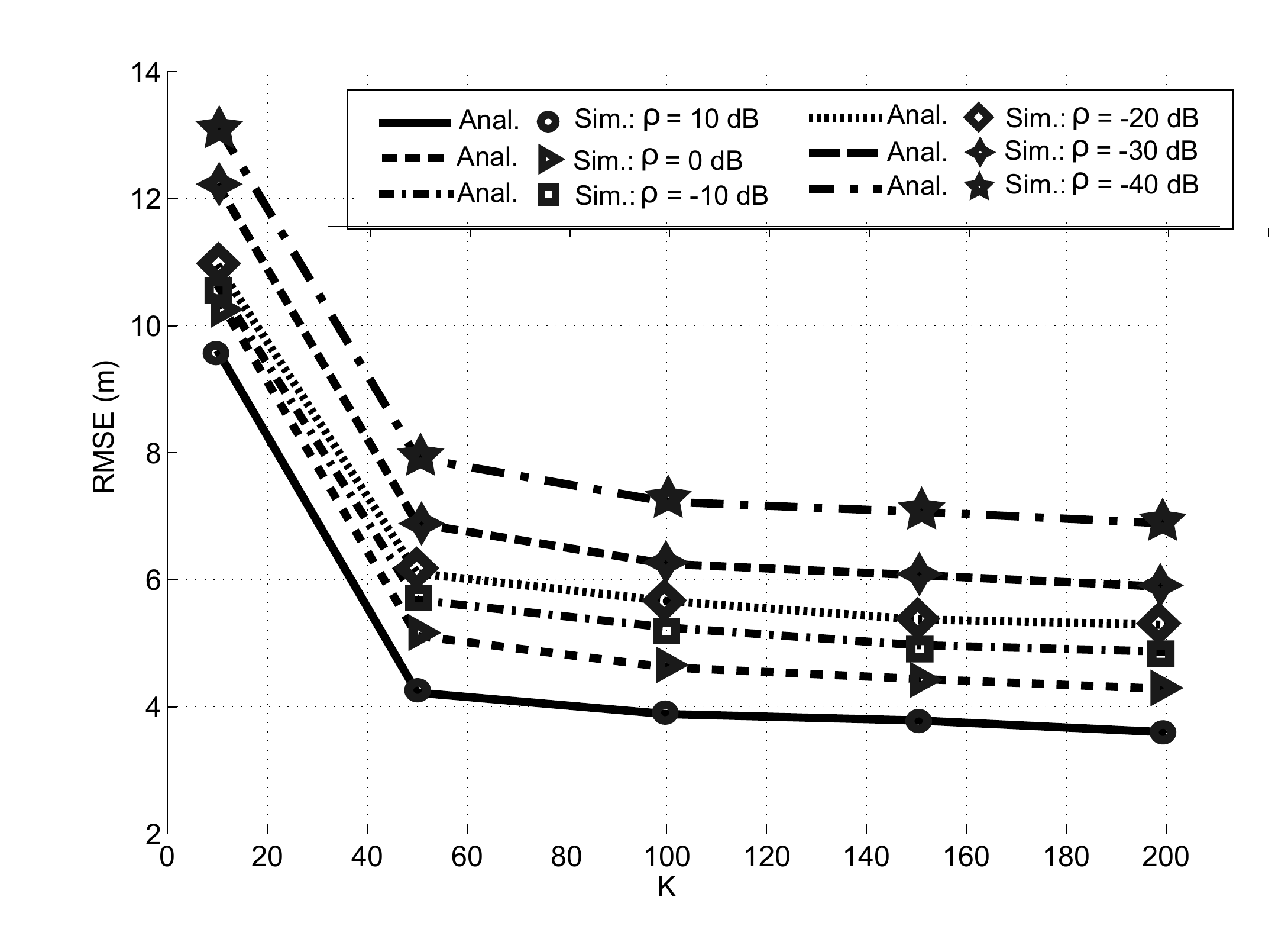}
		\caption{{Improved Cyclic WCL}}
		\label{fig:diff_K_icwcl}
	\end{subfigure}
	\caption{Impact of CR density on a) Cyclic WCL and b) Improved Cyclic WCL, with $p_t=10$dBm. Number of samples $N=500$. Target and interferer locations: $[0, 0]$, $[20, 20]$.}
	\label{fig:rmse_vs_rho_diff_K}
\end{figure}

\subsubsection{Impact of CR density}
The impact of increased CR density in the network on the localization error is shown in Fig. \ref{fig:rmse_vs_rho_diff_K}. It has been observed that increasing the number of CRs from $10$ to $200$ at $\rho=10$dB reduces the error by $9$m, as shown in Fig. \ref{fig:diff_K_cwcl}. However, at $\rho=-40$ dB increasing the number of CRs reduces the error by only $0.75$m. This is due to the fact that increasing $K$ increases the number of CRs in the vicinity of the interferer as well which contribute to increased error at higher interferer power ($\rho=-40$dB). Therefore, any gain obtained by increasing the CR density is compensated by increased interferer power which results in essentially a flat curve at $\rho = -40$dB. {Further, the localization error in improved Cyclic WCL, as shown in Fig.} \ref{fig:diff_K_icwcl}, {decreases by approximately $5$m  if $K$ is increased from $10$ to $50$ for all values of $\rho$, while the error is reduced by approximately $1$m if $K$ is increased from $50$ to $200$. Therefore, the performance of the proposed algorithm changes only by a small amount if $K \geq 50$, irrespective of the interferer power.}

\subsubsection{Impact of number of samples N}
\begin{figure}[t!]
	\centering
	\begin{subfigure}[c]{0.6\textwidth}
		\includegraphics[width=\textwidth]{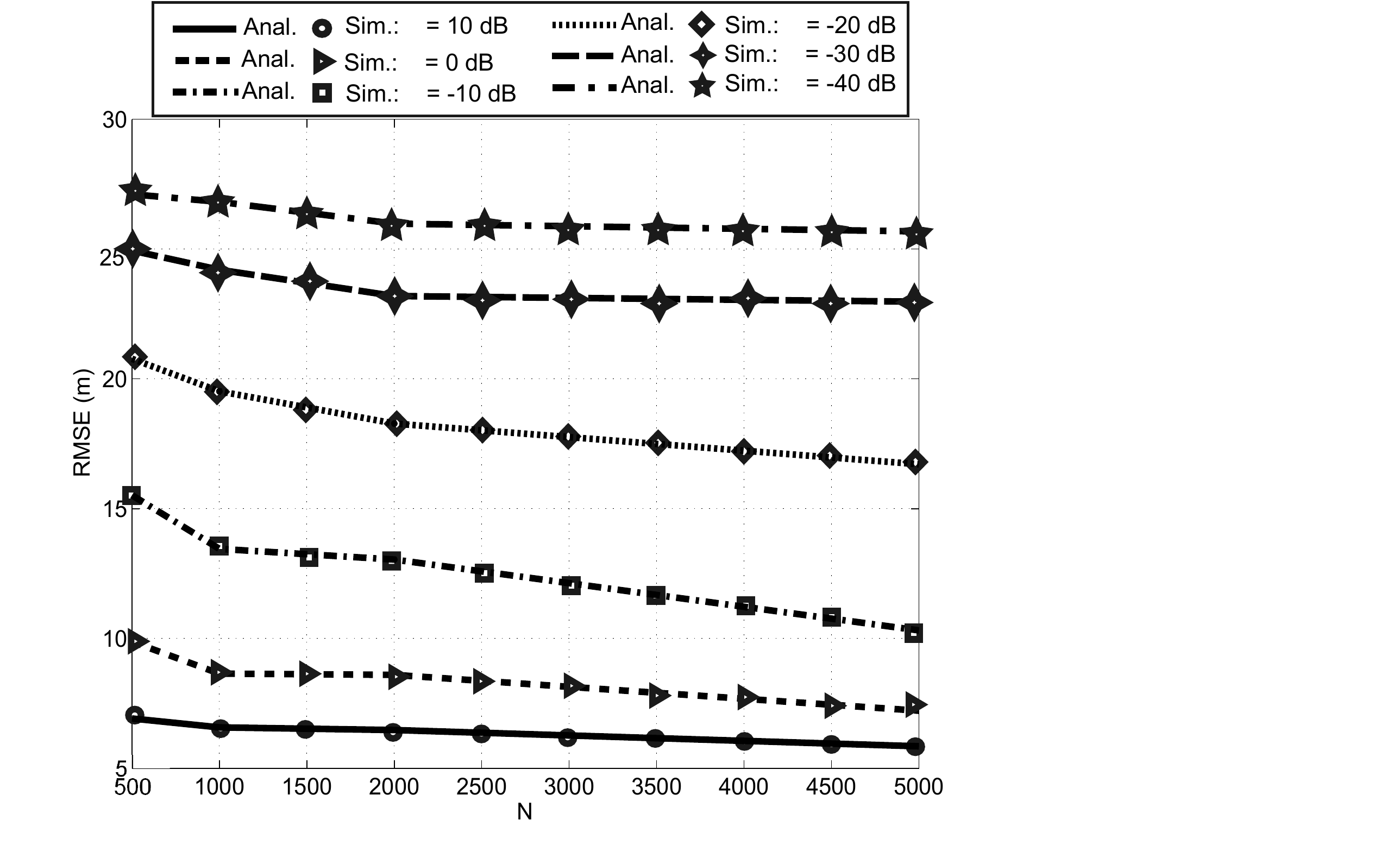}
		\caption{Cyclic WCL}
		\label{fig:diff_N_cwcl}
	\end{subfigure}
	\begin{subfigure}[c]{0.45\textwidth}
		\includegraphics[width=\textwidth]{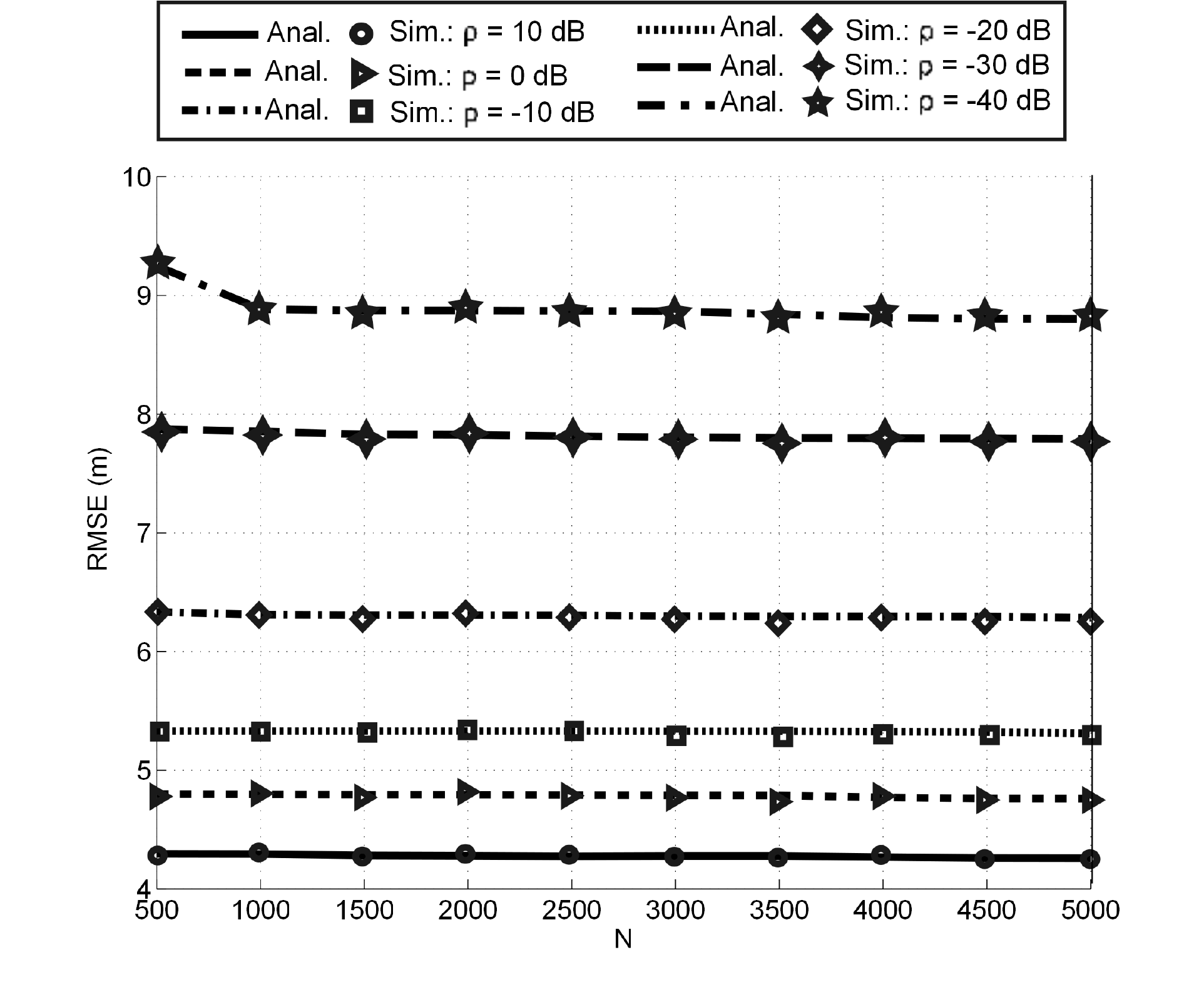}
		\caption{{Improved Cyclic WCL}}
		\label{fig:diff_N_icwcl}
	\end{subfigure}
	\caption{Impact of number of samples ($N$) on a) Cyclic WCL and b) Improved Cyclic WCL, with $p_t=10$dBm. Number of CRs $K=50$. Target and interferer locations: $[0, 0]$, $[20, 20]$.}
	\label{fig:rmse_vs_rho_diff_N}
\end{figure}

In the Cyclic WCL, non-asymptotic estimate of the CAC of the received signal (\ref{eq:cac_rk}) is used to compute weights for each CR location. The non-asymptotic estimates are based on $N$ samples of the received signal $r_k$. Therefore, performance of the Cyclic WCL and the improved Cyclic WCL depends on the value of $N$.
The impact of the value of $N$ on the performance of the  algorithm is shown in Fig. \ref{fig:rmse_vs_rho_diff_N}. It is observed that increasing the number of samples reduces the error in Cyclic WCL. For example, the error is reduced by up to $5$m when $N$ is increased from 500 to 5000. With increased value of $N$, the estimate $\hat{R}_{r_k}^{\alpha_t}$ approaches the true value of $R_{r_k}^{\alpha_t}$ resulting in lower value of the error. {On the other hand, the error in the improved Cyclic WCL does not change significantly with increased $N$ as shown in Fig.} \ref{fig:diff_N_icwcl}. {Therefore, the performance of improved Cyclic WCL is independent of number of samples $N$ as long as $N> \lceil \frac{f_s}{\Delta\hat{\alpha}}\rceil$.}

\subsection{Impact of imperfect knowledge of $\alpha_i$}
\label{sec:Results_Delta_Alpha}
\begin{figure}[t!]
	\centering
	\includegraphics[width=0.9\columnwidth]
	{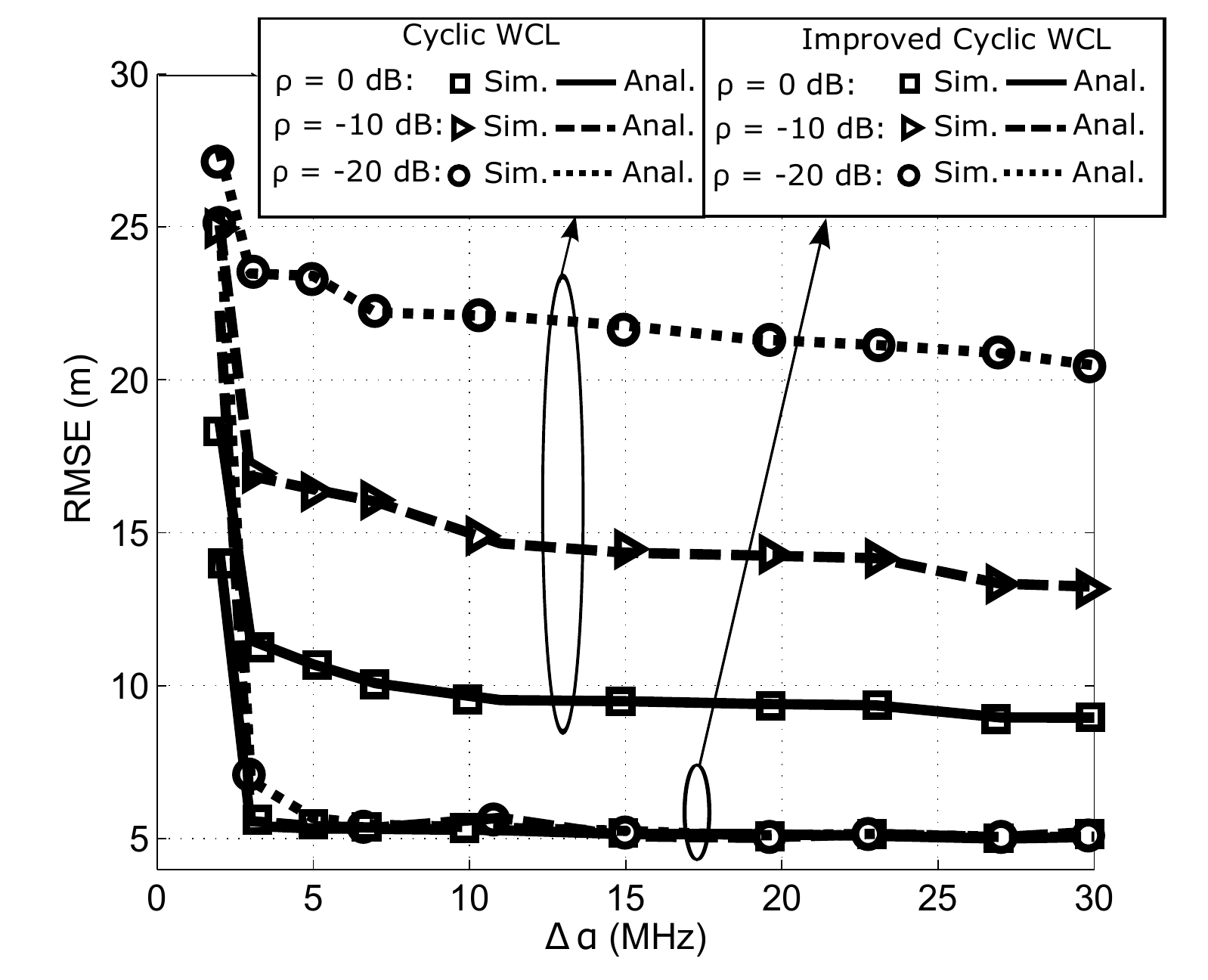}
	\caption{RMSE vs $\Delta\alpha = |\alpha_t - \alpha_i|$ with $\alpha_t = 20$ MHz, Estimated cyclic frequency of interferer $\hat{\alpha}_i = 25$ MHz. Estimated difference in cyclic frequencies: $\Delta \hat{\alpha} = |\hat{\alpha}_i - \alpha_t| = 5$ MHz. Number of samples $N=500>10 \lceil \frac{f_s}{\Delta\hat{\alpha}}\rceil$. Target and interferer locations: $[0, 0]$, $[20, 20]$. Suboptimal threshold ${\phi}_0^{sub}$ is used in improved Cyclic WCL.}
	\label{fig:delta_alpha}
\end{figure}

In both Cyclic WCL and improved Cyclic WCL, the first step is to select the number of samples $N$ based on $\Delta \alpha = |\alpha_t - \alpha_i|$. In this section, the cyclic frequency of the interferer $\alpha_i$ is varied from $22$MHz to $50$MHz. In order to compute $N$, we assume $\hat{\alpha}_i=25$MHz. The number of samples used are $N=500$, which satisfies $N>10 \lceil \frac{f_s}{\Delta\hat{\alpha}}\rceil =  400$, where $\Delta \hat{\alpha} = |\hat{\alpha}_i - \alpha_t| = 5$MHz. The impact of the imperfect knowledge of $\Delta \alpha$ is shown in Fig. \ref{fig:delta_alpha}. If $\Delta \alpha < \Delta \hat{\alpha} = 5$MHz, the number of samples do not satisfy the condition $N>10 \lceil \frac{f_s}{\Delta\hat{\alpha}}\rceil$, which results in higher interference component in the CAC. This phenomenon results in higher error in both Cyclic WCL and improved Cyclic WCL. On the other hand, if $\Delta \alpha > \Delta \hat{\alpha} = 5$MHz, then the error is reduced and the improvement in the performance of the Cyclic WCL depends on the transmit power ratio ($\rho$). Further, it can be observed that the performance of the improved Cyclic WCL remains the same for different $\rho$ and $\Delta \alpha$ in the regime $\Delta \alpha > \Delta \hat{\alpha}=5$MHz. Therefore, the improved Cyclic WCL is robust to both interference power and error in the estimation of $\alpha_i$.

\subsection{Comparison between traditional WCL, Cyclic WCL and improved Cyclic WCL}
\label{sec:Results_Comparison}
\begin{figure}[t!]
	\centering
	\includegraphics[width=\columnwidth]
	{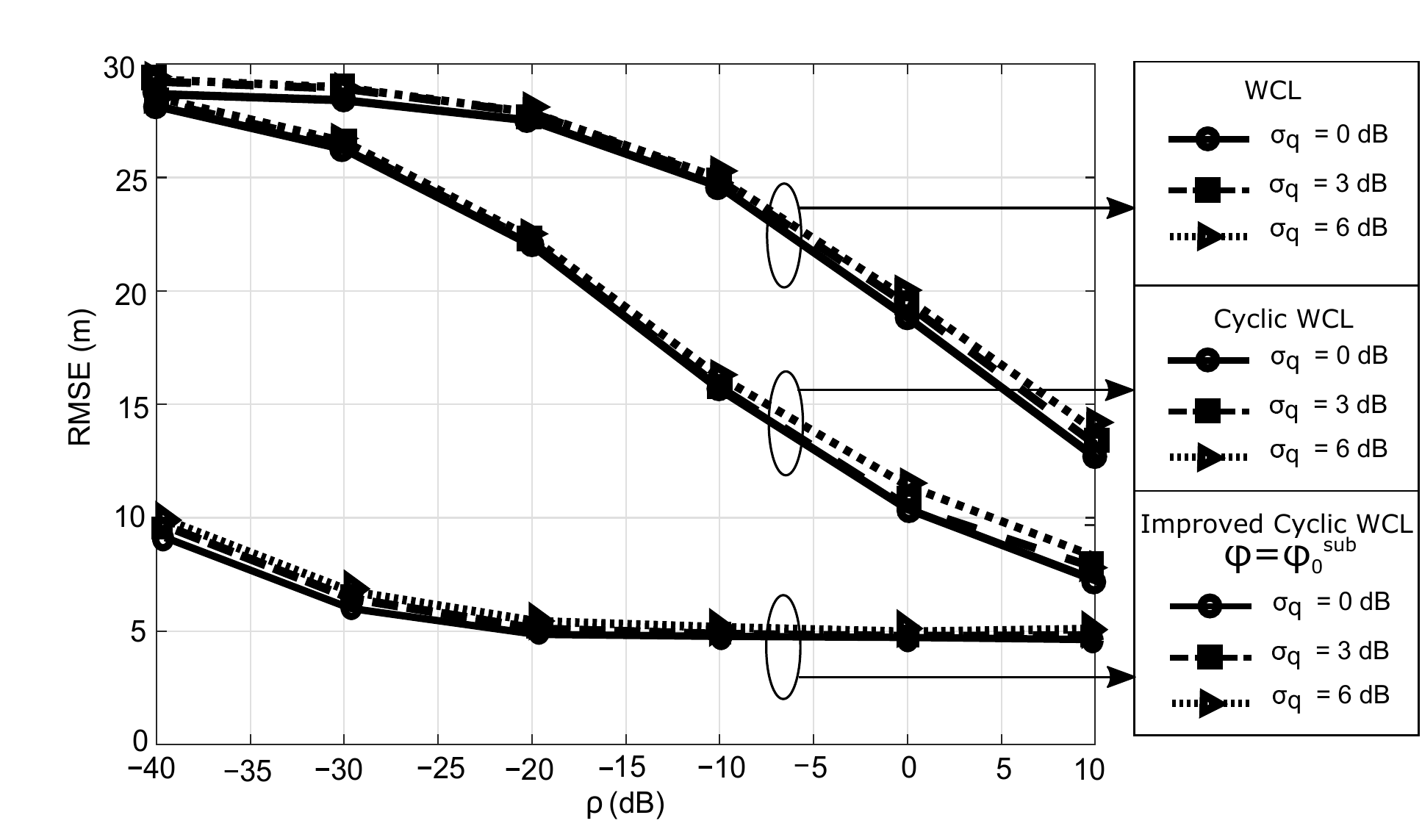}
	\caption{Impact of shadowing on WCL, Cyclic WCL, and improved Cyclic WCL with suboptimal FVC threshold. Number of CRs $K = 50$. Number of samples $N=500$. Target and interferer locations: $[0, 0]$, $[20, 20]$.}
	\label{fig:shadowing}
\end{figure}

In this section, we compare the performances of the traditional WCL, Cyclic WCL and the improved Cyclic WCL in shadowing environment. It is observed that, even in shadowing environment ($\sigma_q = 6$ dB), the error in Cyclic WCL is smaller than traditional WCL, as shown in Fig. \ref{fig:shadowing}. In the case of improved Cyclic WCL, the error is reduced by a factor of three as compared to the traditional WCL for $\rho=-40$dB. Further, it has been observed that Cyclic WCL algorithms are robust to shadowing. For example, in improved Cyclic WCL, the error has increased by only $2$m when shadowing variance has increased four fold from $\sigma_q = 0$dB to $\sigma_q=6$dB. This is due to the fact that the shadowing effect over $K$ CRs averages out in the WCL algorithms.

\begin{figure}[t!]
	\centering
	\includegraphics[width= \columnwidth]{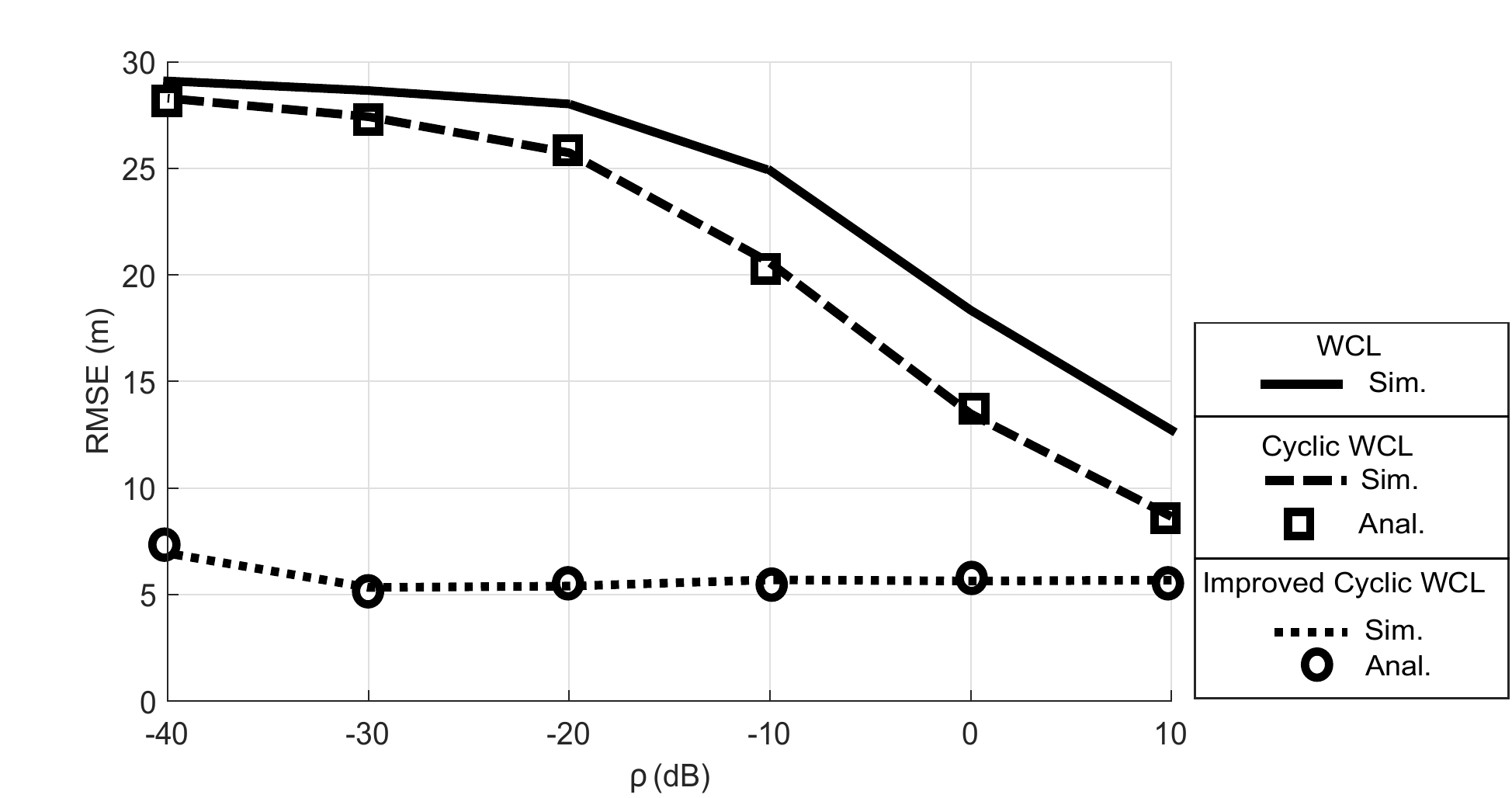}
	\caption{ RMSE in WCL, Cyclic WCL and improved Cyclic WCL algorithms with OFDM signals. Target and interferer cyclic frequencies $\alpha_t = 250$kHz and $\alpha_i=14$kHz, respectively. Sampling frequency at CRs $f_s=500$kHz. Number of samples $N=100$.  Number of CRs $K=50$. Target and interferer locations: $[0,0]$, $[20,20]$.}
	\label{fig:ofdm_signals}
\end{figure}

Next, we show the performance of proposed algorithms with OFDM signals. In these results, $s_t$ is a WLAN signal with 64 sub-carriers,  and sub-carrier spacing $\Delta f_t =312.5$kHz. The OFDM symbol duration $T_g=4 \mu$s with $T_{cp}=0.8\mu$s. The cyclic frequency of $s_t$ is $\alpha_t=1/T_g=250$kHz.  The interference, $s_i$, is a LTE signal with 1024 sub-carriers and sub-carrier spacing of $\Delta f_i=15$kHz. The OFDM symbol duration $T_h=71.4 \mu$s with $T_{cp}=4.7 \mu$s. The cyclic frequency of the interference is $\alpha_i=1/T_h=14$kHz. Both target and interfering signals carry 4-QAM signal on each sub-carrier, i.e., $c_{l,k}$ and $d_{l,k}$ are 4-QAM symbols. The sampling frequency $f_s = 500$kHz, while the number of samples $N=100>10 \lceil \frac{f_s}{\Delta \alpha} \rceil$. Localization errors in the traditional WCL, Cyclic WCL, and improved Cyclic WCL are shown in Fig. \ref{fig:ofdm_signals}.  The figure also shows analytically computed RMSE with OFDM signals following the derivations in Appendix B. It has been observed that the proposed improved Cyclic WCL continues to provide three times lower error as compared to traditional WCL for $\rho=-40$dB. Further, the comparison of RMSE in improved Cyclic WCL in Fig. \ref{fig:shadowing} and \ref{fig:ofdm_signals} shows that the proposed algorithm performs equally well under OFDM and single carrier signals.

\subsection{Performance under multipath fading channels}
\label{sec:Results_multipath}
The three algorithms are also studied under multipath fading channels suitable for indoor and outdoor settings. For indoor setting, TGn channel model from WLAN standard is used \cite{TGn}. This channel model has delay spread of $40$ns and Doppler spread of $10$Hz. For outdoor setting, extended typical urban (ETU) channel model is considered with $5\mu$s delay spread  and  $300$Hz Doppler spread \cite{LTE_36_101}. This is a commonly used model in LTE cellular system. The channel coefficients $h_{t,k}$ and $h_{i,k}$ are obtained from the aforementioned channel models, where $h_{t,k}$ is the channel between the target and the $k^{th}$ CR, while  $h_{i,k}$ is the channel between interferer and $k^{th}$ CR. The received signal is then obtained as $r_k=\sqrt{p_{t,k}}\left[h_{t,k} \ast s_t \right]+\sqrt{p_{i,k}} \left[h_{i,k} \ast s_i \right]+ w_k$, where $\ast$  denotes the convolution operation. Further implementations of the three algorithms remains the same as in AWGN channel.

\begin{figure}[t!]
	\centering
	\includegraphics[width= \columnwidth]{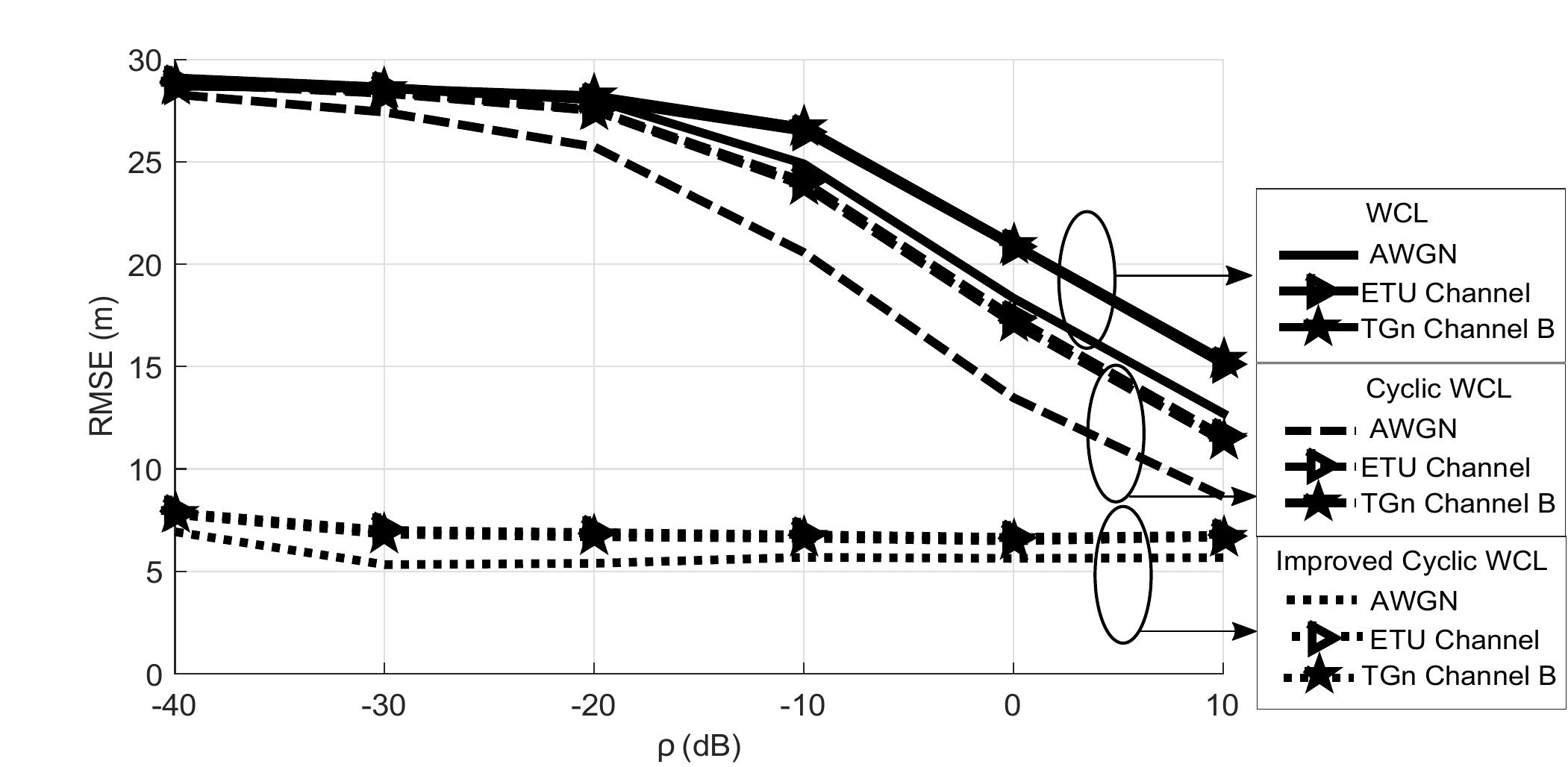}
	\caption{RMSE in WCL, Cyclic WCL, and improved Cyclic WCL with different channel models. Signal parameters are same as in Fig. \ref{fig:ofdm_signals}.	}
	\label{fig:nlos_channels}
\end{figure}

The localization errors under the two multipath channels are shown in Fig. \ref{fig:nlos_channels}. As in AWGN case $N=100$ samples are used at sampling rate $500$kHz for localization. It has been observed that the localization errors in all three algorithms increase as compared to AWGN channel. This increase in error can be explained as follows. The signal observation interval at each CR is $100/500$kHz $=0.2$ms, which is smaller than the coherence time of TGn and ETU channels. Therefore, the received power at each CR in this duration is affected by the small scale fading and the impact of fading is not averaged out as the observation duration is smaller than the coherence time. This leads to increased localization errors. However, we can observe that the localization error in WCL and Cyclic WCL is increased by up to $4$m. On the other hand, the error in improved Cyclic WCL increases by up to $2$m only. Therefore, we can conclude that the proposed improved Cyclic WCL more robust against multipath fading channels and continues to provide significant performance gain over the WCL algorithm. 

\section{Conclusion}
\label{sec:Conclusion}
In this paper, we have proposed the Cyclic WCL algorithm to mitigate the adverse impact of a spectrally overlapped interference in CR networks on the target localization process. The proposed algorithm uses cyclic autocorrelation of the received signal in order to estimate the target location. Theoretical analysis of the  algorithm is presented in order to compute the error in localization. We have also proposed the improved Cyclic WCL algorithm that identifies CRs in the vicinity of interferer and eliminates them from the localization process to further reduce the error. 

We have studied impacts of interferer's power, its location, and CR density on performance of the proposed algorithms. The improved Cyclic WCL is observed to be robust against interferer's location and its transmit power. The comparison between the traditional WCL and the improved Cyclic WCL shows that the proposed algorithm provides significantly lower error in localization when there is a spectrally overlapped interference in the network. It has been observed that the improved Cyclic WCL is also robust against shadowing and multipath fading environment. The proposed localization algorithm performs equally well with single carrier as well as OFDM signals.


\appendices

\section{Proof: $\hat x_t$ is a ratio of quadratic form in $\boldsymbol{\hat \theta}$}
\label{app_rq}
From (\ref{eq:xt}), we have 
\begin{align}
\nonumber
{{\hat x}_t} &= \frac{{\sum\limits_{k = 1}^K {|{{\hat R}_{{r_k}}^{\alpha_t}}{|^2}} {x_k}}}
{{\sum\limits_{k = 1}^K {|{{\hat R}_{{r_k}}^{\alpha_t}}{|^2}} }} = \frac{{{{\sum\limits_{k = 1}^K {\left\| {\left[ \begin{gathered}
  {\boldsymbol{\hat \theta_r }^T} \hfill \\
  {\boldsymbol{\hat \theta_i }^T} \hfill \\ 
\end{gathered}  \right]\mathbf{p_k}} \right\|} }^2}{x_k}}}
{{\sum\limits_{k = 1}^K {{{\left\| {\left[ \begin{gathered}
  {\boldsymbol{\hat \theta_r }^T} \hfill \\
  {\boldsymbol{\hat \theta_i }^T} \hfill \\ 
\end{gathered}  \right]\mathbf{p_k}} \right\|}^2}} }} \text{ and}
\end{align}
\begin{align}
\hat{x}_t= \frac{{\sum\limits_{k = 1}^K {\mathbf{p_k}^T\left[ {{\boldsymbol{\hat \theta_r }}{\text{ }}{\boldsymbol{\hat \theta_i }}} \right]} {\text{ }}{x_k}\left[ \begin{gathered}
  {\boldsymbol{\hat \theta_r }^T} \hfill \\
  {\boldsymbol{\hat \theta_i }^T} \hfill \\ 
\end{gathered}  \right]\mathbf{p_k}}}
{{\sum\limits_{k = 1}^K {\mathbf{p_k}^T\left[ {{\boldsymbol{\hat \theta_r }}{\text{ }}{\boldsymbol{\hat \theta_i }}} \right]} {\text{ }}\left[ \begin{gathered}
  {\boldsymbol{\hat \theta_r }^T} \hfill \\
  {\boldsymbol{\hat \theta_i }^T} \hfill \\ 
\end{gathered}  \right]\mathbf{p_k}}}.
\end{align}
After rearranging the terms and taking the summation operator inside, we have
\begin{align}
\nonumber
{{\hat x}_t} = \frac{{Tr\left( {\left[ \begin{gathered}
  {\boldsymbol{\hat \theta_r }^T} \hfill \\
  {\boldsymbol{\hat \theta_i }^T} \hfill \\ 
\end{gathered}  \right]\mathop {\sum {\text{ }}}\limits_{k = 1}^K \mathbf{p_k}{\text{ }}{x_k}\mathbf{p_k}^T\left[ {{\boldsymbol{\hat \theta_r }}{\text{ }}{\boldsymbol{\hat \theta_i }}} \right]} \right)}}
{{Tr\left( {\left[ \begin{gathered}
  {\boldsymbol{\hat \theta_r }^T} \hfill \\
  {\boldsymbol{\hat \theta_i }^T} \hfill \\ 
\end{gathered}  \right]\mathop \sum \limits_{k = 1}^K  \mathbf{p_k}\mathbf{p_k}^T\left[  {{\boldsymbol{\hat \theta_r }}{\text{ }}{\boldsymbol{\hat \theta_i }}} \right]} \right)}}.
\end{align}
Further, using the definitions of $\boldsymbol{P}$ and $\boldsymbol{X}$ from (\ref{eq:matrix_def}), we get
\begin{align}
\nonumber
{{\hat x}_t} &= \frac{{Tr\left( {\left[ \begin{gathered}
  {\boldsymbol{\hat \theta_r }^T} \hfill \\
  {\boldsymbol{\hat \theta_i }^T} \hfill \\ 
\end{gathered}  \right]\mathbf{PXP}^T\left[  {{\boldsymbol{\hat \theta_r }}{\text{ }}{\boldsymbol{\hat \theta_i }}} \right]} \right)}}
{{Tr\left( {\left[ \begin{gathered}
  {\boldsymbol{\hat \theta_r }^T} \hfill \\
  {\boldsymbol{\hat \theta_i }^T} \hfill \\ 
\end{gathered}  \right]\mathbf{PP}^T\left[  {{\boldsymbol{\hat \theta_r }}{\text{ }}{\boldsymbol{\hat \theta_i }}} \right]} \right)}} \\
\nonumber &= \frac{{\left[ {{\boldsymbol{\hat \theta_r }^T}{\text{ }}{\boldsymbol{\hat \theta_i }^T}} \right]\left[ {\begin{array}{*{20}{c}}
   \mathbf{PXP}^T & \boldsymbol{0}  \\
   \boldsymbol{0} & \mathbf{PXP}^T  \\
 \end{array} } \right]\left[ \begin{gathered}
  {\boldsymbol{\hat \theta_r }} \hfill \\
  {\boldsymbol{\hat \theta_i }} \hfill \\ 
\end{gathered}  \right]}}
{{\left[ {{\boldsymbol{\hat \theta_r }^T}{\text{ }}{\boldsymbol{\hat \theta_i }^T}} \right]\left[ {\begin{array}{*{20}{c}}
   \mathbf{PP}^T & \boldsymbol{0}  \\
   \boldsymbol{0} & \mathbf{PP}^T  \\
 \end{array} } \right]\left[ \begin{gathered}
  {\boldsymbol{\hat \theta_r }} \hfill \\
  {\boldsymbol{\hat \theta_i }} \hfill \\ 
\end{gathered}  \right]}}= \frac{{{\boldsymbol{\hat \theta }^T}\mathbf{A_x}\boldsymbol{\hat \theta} }}
{{{\boldsymbol{\hat \theta }^T}\mathbf{B}\boldsymbol{\hat \theta} }},
\end{align}
where $\mathbf{A_x} = \diag(\mathbf{PXP^T, PXP^T})$ and $\mathbf{B} =\diag(\mathbf{PP^T, PP^T}) $.

\section{Derivations of $E[\boldsymbol{\hat \theta}]$ and $\boldsymbol{\Sigma_{\hat \theta}}$ in terms of transmitted symbols and pulse shape.}
\label{app_derivations}
The mean and covariance matrix of $\boldsymbol{\hat{\theta}} = [\boldsymbol{\hat{\theta_r}}^T  \boldsymbol{\hat{\theta_i}}^T ]^T$ are defined
as follows:
\begin{align}
\nonumber
E[\boldsymbol{\hat{\theta}}]=[E[\boldsymbol{\hat{\theta}_{r}}]^{T},E[\boldsymbol{\hat{\theta}_{i}}]^{T}]^{T} \text{, }
\boldsymbol{\Sigma_{\hat{\theta}}}=E[\boldsymbol{\hat{\theta}\hat{\theta}}^{T}]-E[\boldsymbol{\hat{\theta}}]E[\boldsymbol{\hat{\theta}}]^{T},
\end{align}
where $E[\boldsymbol{\hat{\theta}\hat{\theta}}^{T}]=\left[\begin{array}{cc}
E[\boldsymbol{\hat{\theta}_{r}\hat{\theta}_{r}}^{T}] & E[\boldsymbol{\hat{\theta}_{r}\hat{\theta}_{i}}^{T}]\\
E[\boldsymbol{\hat{\theta}_{i}\hat{\theta}_{r}}^{T}] & E[\boldsymbol{\hat{\theta}_{i}\hat{\theta}_{i}}^{T}]
\end{array}\right]$

In order to compute $E[\boldsymbol{\hat{\theta}}]$ and $\boldsymbol{\Sigma_{\hat{\theta}}}$, the terms of the form $E[\operatorname {Re}\{\hat{R}_{u}^{\alpha_t}\}\operatorname {Im}\{\hat{R}_{v}^{\alpha_t}\}]$, $E[\operatorname {Re}\{\hat{R}_{u}^{\alpha_t}\}\operatorname {Im}\{\hat{R}_{uv}^{\alpha_t}\}]$ and $E[\operatorname {Re}\{\hat{R}_{uv}^{\alpha_t}\}\operatorname {Im}\{\hat{R}_{uv}^{\alpha_t}\}]$ are required, where $u,v \in \{s_t, s_i, w_k\}$. Further, it should be noted that the four terms of the form $E[\operatorname {Re}\{\hat{R}_{u}^{\alpha_t}\}\operatorname {Re}\{\hat{R}_{v}^{\alpha_t}\}]$, $E[\operatorname {Im}\{\hat{R}_{u}^{\alpha_t}\}\operatorname {Re}\{\hat{R}_{v}^{\alpha_t}\}]$, $E[\operatorname {Re}\{\hat{R}_{u}^{\alpha_t}\}\operatorname {Im}\{\hat{R}_{v}^{\alpha_t}\}]$ and $E[\operatorname {Im}\{\hat{R}_{u}^{\alpha_t}\}\operatorname {Im}\{\hat{R}_{v}^{\alpha_t}\}]$ can be obtained from a single term $E[\hat{R}_{u}\hat{R}_{v}^*]$, by replacing $e^{-j2\pi\alpha_t(n-m)T_s}$ with $\operatorname{cos}(2\pi\alpha_tnT_s)\operatorname{cos}(2\pi\alpha_tmT_s)$, $\operatorname{sin}(2\pi\alpha_tnT_s)\operatorname{cos}(2\pi\alpha_tmT_s)$, $\operatorname{cos}(2\pi\alpha_tnT_s)\operatorname{sin}(2\pi\alpha_tmT_s)$ and $\operatorname{sin}(2\pi\alpha_tnT_s)\operatorname{sin}(2\pi\alpha_tmT_s)$, respectively. Therefore, only the computations of the form $E[\hat{R}_{u}^{\alpha_t}\hat{R}_{v}^{\alpha_t*}]$ are shown.
\subsection{Moments of CACs and CCCs of signals $s_t$ and $s_i$}

The moments of CACs and CCCs are computed using the moments of $s_t$ and $s_i$, as shown later in this section. Therefore, the moments of $s_t$ and $s_i$ are computed first. 


Since $a_l's$ and $c_{\kappa,l}$ zero mean, we have $E[s_{t}(n)]=0$ under both single- and multi-carrier models. In single carrier signal $s_t$, the second moments of $s_t$ are
\begin{align}
\small
\nonumber
E[s_{t}(n)s_{t}(m)]&=\sum_{k=-\infty}^{\infty}\sum_{l=-\infty}^{\infty}E[a_{k}a_{l}]g_{n,k}g_{m,l}e^{j2\pi f_t(n+m)T_s}\\\nonumber
&=E[a^{2}]e^{j2\pi f_t(n+m)T_s}\sum_{k=-\infty}^{\infty}g_{n,k}g_{m,k} \text{, and}
\\
E[s_{t}(n)s_{t}^{*}(m)]&=E[|a|^{2}]e^{j2\pi f_t(n-m)T_s}\sum_{k=-\infty}^{\infty}g_{n,k}g_{m,k}.
\label{eq:app1}
\end{align}

Further, the subscript $l$ is dropped after taking the expectation in the derivations, since $a_l's$ and $c_{\kappa,l}$ are i.i.d. for all $l$.
In order to compute $E[|s_{t}(n)|^{2}|s_{t}(m)|^{2}]$ for single carrier signals, note that $|s_{t}(n)|^{2}= \sum_{k=-\infty}^{\infty}\sum_{l=-\infty}^{\infty}a_{k}a_{l}^{*}g_{n,k}g_{n,l}$ and $|s_{t}(m)|^{2}=\sum_{p=-\infty}^{\infty}\sum_{q=-\infty}^{\infty}a_{p}a_{q}^{*}g_{m,p}g_{m,q}$. Then, we have
\begin{multline}
\nonumber E[|s_{t}(n)|^{2}|s_{t}(m)|^{2}]=\\\sum_{k=-\infty}^{\infty}\sum_{l=-\infty}^{\infty}\sum_{p=-\infty}^{\infty}\sum_{q=-\infty}^{\infty}E[a_{k}a_{l}^{*}a_{p}a_{q}^{*}]g_{n,k}g_{n,l}g_{m,p}g_{m,q}.
\end{multline}
Since $a_k's$ are i.i.d. and zero mean, we have:
\begin{multline}
E[|s_{t}(n)|^{2}|s_{t}(m)|^{2}] = E[|a|^{4}]\sum_{k=-\infty}^{\infty}g^{2}_{n,k}g^{2}_{m,k} 
\\ + E[|a|^{2}]^{2}\sum_{k=-\infty}^{\infty}\sum_{p=-\infty, p \neq k}^{\infty}g^{2}_{n,k}g^{2}_{m,p} 
\\+|E[a^{2}]|^{2}\sum_{k=-\infty}^{\infty}\sum_{l=-\infty, l \neq k}^{\infty}g_{n,k}g_{n,l}g_{m,k}g_{m,l}
\\+E[|a|^{2}]^{2}\sum_{k=-\infty}^{\infty}\sum_{l=-\infty, l \neq k}^{\infty}g_{n,k}g_{n,l}g_{m,l}g_{m,k}.
\label{eq:app2}
\end{multline}

 For the OFDM signal, $E[s_{t}(n)s_{t}(m)]$, $E[s_{t}(n)s_{t}^{*}(m)]$, and $E[|s_{t}(n)|^{2}|s_{t}(m)|^{2}]$ are obtained by making the following substitutions in (\ref{eq:app1}) and (\ref{eq:app2}). First, $E[a^2]$ is substituted with  $\sum_{\kappa \in \mathbb{K}}E[c_{\kappa}^2]e^{j2\pi \kappa \Delta f_t (n+m)T_s}$, $E[|a|^2]$ is substituted with $\sum_{\kappa \in \mathbb{K}}E[|c_{\kappa}|^2]e^{j2\pi \kappa \Delta f_t (n-m)T_s}$, where $\mathbb{K} = \left \{- \frac{N_{c,t}}{2},..., \frac{N_{c,t}}{2}-1 \right\}$. Finally, $E[|a|^4]$ is substituted with 
\begin{align}
\small
\nonumber \sum_{\kappa \in \mathbb{K}} E[|c_{\kappa}|^4] + \sum_{\kappa_1 \in \mathbb{K}} \sum\limits_{\substack{\kappa_2 \in \mathbb{K}\\ \kappa_1 \neq \kappa_2}}E[c_{\kappa_1}^2]E[(c^{*}_{\kappa_2})^2]e^{j 2 \pi (\kappa_1 - \kappa_2)\Delta f_t(n+m)T_s } 
\\+\sum_{\kappa_1 \in \mathbb{K}} \sum\limits_{\substack{\kappa_2 \in \mathbb{K}\\ \kappa_1 \neq \kappa_2}}E[|c_{\kappa_1}|^2]E[|c_{\kappa_2}|^2]\left(1+ e^{j 2 \pi (\kappa_1 - \kappa_2)\Delta f_t(n-m)T_s }\right).
\label{eq:app2_2}
\end{align}



Similarly, the moments are computed for $s_{i}(n)$ using single carrier and OFDM signal models.
From (\ref{eq:app1}) and (\ref{eq:app2_2}), first and second order moments of the estimates of the CAC are obtained as follows:
\begin{align}
\nonumber
E[ \hat{R}_{s_{t}}^{\alpha_t}]&=\frac{1}{N}\sum_{n=0}^{N-1}E[|s_{t}(n)|^{2}]e^{-j2\pi \alpha_t n T_s} \text{, and} \\
E[|\hat{R}_{s_{t}}^{\alpha_t}|^2]&=\frac{1}{N^{2}}\sum_{n=0}^{N-1}\sum_{m=0}^{N-1}E[|s_{t}(n)|^{2}|s_{t}(m)|^{2}]e^{-j2\pi \alpha_t (n-m) T_s}.
\label{eq:app3}
\end{align}
 Further, the cross correlation between estimates of CACs of $s_t$ and $s_i$ is
\begin{multline}
\small E[\hat{R}_{s_{t}}^{\alpha_t}\hat{R}_{s_{i}}^{\alpha_t*}]=\\\frac{1}{N^{2}}\sum_{n=0}^{N-1}\sum_{m=0}^{N-1}E[|s_{t}(n)|^{2}]E[|s_{i}(m)|^{2}]e^{-j2\pi \alpha_t (n-m) T_s}.
\end{multline}
Similarly, the mean of the CCCs between $s_t$ and $s_i$ is
\begin{align}
\nonumber
\small E[\hat{R}_{s_{t}s_{i}}^{\alpha_t}]=&E \left[\frac{1}{N}\sum_{n=0}^{N-1}2\operatorname {Re}\{s_{t}(n)s_{i}^{*}(n\}e^{-j2\pi\alpha nT_{s}}\right]\\\nonumber=&\frac{1}{N}\sum_{n=0}^{N-1}(E[s_{t}(n)]E[s_{i}(n)]^{*}\\&+E[s_{t}(n)]^{*}E[s_{i}(n)])e^{-j2\pi \alpha_t n T_s}=0.
\label{eq:app5}
\end{align}
The above equation follows from the fact that $s_t$ and $s_i$ are independent and
zero mean signals. From the same argument, we can show that $E[\hat{R}_{s_{t}s_{i}}^{\alpha_t}\hat{R}_{s_{i}}^{\alpha_t*}]=0$ and $E[\hat{R}_{s_{t}s_{i}}\hat{R}_{s_{t}}^*]=0$.

Finally, in order to compute the second moment of CCC $E[\hat{R}_{s_{t}s_{i}}\hat{R}_{s_{t}s_{i}}^*]$, let is define $Z$ as follows:
\begin{align}
\nonumber
Z=&E[2\operatorname {Re}\{s_{t}(n)s_{i}^{*}(n)\}2\operatorname {Re}\{s_{t}(m)s_{i}^{*}(m)\}] \\\nonumber= &2\operatorname {Re}\{E[s_{t}(n)s_{t}(m)]E[s_{i}(n)s_{i}(m)]^{*}\}\\&+2\operatorname {Re}\{E[s_{t}(n)s_{t}^{*}(m)]E[s_{i}^{*}(n)s_{i}(m)]\}.
\label{eq:app7}
\end{align}
Then, it can be shown that
\begin{align}
E[\hat{R}_{s_{t}s_{i}}^{\alpha_t}\hat{R}_{s_{t}s_{i}}^{\alpha_t*}]=\frac{1}{N^{2}}\sum_{n=0}^{N-1}\sum_{m=0}^{N-1}Ze^{-j2\pi \alpha_t (n-m)T_s}.
\label{eq:app8}
\end{align}
In the equations (\ref{eq:app3})-(\ref{eq:app8}), the required moments of $s_t$ and $s_i$ are obtained using (\ref{eq:app1}) and (\ref{eq:app2}).
The moments related to the CAC of the noise are obtained using (\ref{eq:app3})-(\ref{eq:app7}) by substituting $s_t$ and $s_i$ by $w$ and are also derived in \cite{Rebeiz2013}. The moments are as follows:
\begin{align}
\nonumber
E[\hat{R}_{w}^{\alpha_t}] = \frac{\sigma_{w}^{2}}{N}e^{-j\pi\alpha_t(N-1)T_{s}}\frac{sin(\pi\alpha_t NT_{s})}{sin(\pi\alpha_t T_{s})}, \\
E[\hat{R}_{w}^{\alpha_t}\hat{R}_{w}^{\alpha_t*}]=\frac{\sigma_{w}^{4}}{N^{2}}+\frac{\sigma_{w}^{4}}{N^{2}}\left(\frac{sin^{2}(\pi\alpha_t NT_{s})}{sin^{2}(\pi\alpha_t T_{s})}\right).
\label{eq:app10}
\end{align}

In order to compute $E[\hat{R}_{s_{t}w}\hat{R}_{s_{t}w}^*]$, we substitute $s_i$ by $w$ in (\ref{eq:app8}) and use 
\begin{align}
\nonumber
Y=&2\operatorname {Re}\{E[s_{t}(n)s_{t}(m)]E[w(n)w(m)]^{*}\}\\&+2\operatorname {Re}\{E[s_{t}(n)s_{t}^{*}(m)]E[w^{*}(n)w(m)]\}.
\end{align}
Then we have, 
\begin{align}
\small E[\hat{R}_{s_{t}w}^{\alpha_t}\hat{R}_{s_{t}w}^{\alpha_t*}]=\frac{2\sigma_{w}^{2}}{N^{2}}\sum_{n=0}^{N-1}E[|s_{t}(n)|^{2}] \approx  \frac{2\sigma_{w}^{2}}{N}.
\label{eq:app11}
\end{align}
The above equation follows from the fact that, for $m\neq n$, $Y=0$, and for $m=n$, $Y=2\operatorname {Re}\{E[|s_{t}(n)|^{2}]E[|w(n)|^{2}]\}=2\sigma_{w}^{2}E[|s_{t}(n)|^{2}]$, since $E[w^{2}(n)]=0$. The approximation holds since $s_t$ is a signal with unit average energy. Similarly we can show that  $E[\hat{R}_{s_{i}w}^{\alpha_t}\hat{R}_{s_{i}w}^{\alpha_t*}]\approx  \frac{2\sigma_{w}^{2}}{N}$. Using (\ref{eq:app1}) - (\ref{eq:app11}), the mean $E[\boldsymbol{\hat{\theta}}]$ and covariance matrix $\boldsymbol{\Sigma_{\hat{\theta}}}$ are computed.
\section{Derivation of lower bound on $N$}
In this section, we further analyze the behavior of the CAC of the interferer signal at cyclic frequency $\alpha_t$ as a function of $N$. From derivations in Appendix B, we have
\begin{align}
\nonumber
E[\hat{R}_{s_{i}}^{\alpha_t}]&=\frac{E[|b|^{2}]}{N}\sum_{n=0}^{N-1}\sum_{k=-\infty}^{\infty}h^{2}_{n,k}e^{-j2\pi\alpha_{t}nT_{s}}\\&=\frac{E[|b|^{2}]}{N}\sum_{n=0}^{N-1}\sum_{k=-\infty}^{\infty}h^{2}(nT_{s}-kT_{h})e^{-j2\pi\alpha_{t}nT_{s}}.
\label{eq:exp_R_si}
\end{align}

Let us define $h'(t) = \sum_{k=-\infty}^{\infty}h^2(t-kT_h)$ and let $H'(\alpha)$ be the Fourier transform of $h'(t)$ at frequency $\alpha$. The period of $h'(t)$ is $T_h$ and its Fourier spectrum consists of spectral lines at frequencies $l/T_h, l=1,2,3,...$. $h'(t)$ can be expressed in terms of Fourier series coefficients $H'(l/T_h)$ as $h'(t)=\sum_{l=-\infty}^{\infty}H'(l/T_h)e^{j2\pi \left(\frac{l}{T_h}\right)t}$. The magnitude of the first spectral line at $1/T_h$ is greater than other spectral lines, since $T_h$ is the fundamental period of $h'(t)$. The first spectral line is located at at the symbol rate of the interference signal $\alpha_i=1/T_h$. Further, $h'(nT_s) =  \sum_{k=-\infty}^{\infty}h^2(nT_s-kT_h)$ is obtained by sampling $h'(t)$ at sampling rate $1/T_s > \text{max}(\alpha_t,\alpha_i)$. Let us define windowed discrete time Fourier transform of $h'(nT_s)$ at frequency $\alpha$ as $H'(\alpha, N, T_s) = \frac{1}{N} \sum_{n=0}^{N-1}h'(nT_s)e^{-j2\pi\alpha nT_s}$, where $N$ is the length of the rectangular window. It should be noted that $f_s = 1/T_s >\text{max}(\alpha_t,\alpha_i)$ to avoid aliased components in $H'(\alpha, N, T_s)$. Thus, (\ref{eq:exp_R_si}) can be expressed as $E[\hat{R}_{s_{i}}^{\alpha_t}]=E[|b|^{2}] H'(\alpha_t, N, T_s)$.

In order to find a condition on $N$ such that the interference caused by $h'(nT_s)$ at $\alpha_t$ is negligible, we consider only the first spectral line located at $\alpha_i=1/T_h$. If the power of this spectral component is reduced to a negligible amount at $\alpha_t$, the remaining spectral components (at $l/T_h, l>1$) will have even less power $\alpha_t$ and will not cause significant interference. Then, (\ref{eq:exp_R_si}) can be written as:
\begin{align}
\nonumber
E[\hat{R}_{s_{i}}^{\alpha_t}]&={E[|b|^{2}]}H'(\alpha_t, N, T_s) \\\nonumber&= \frac{E[|b|^{2}]}{N} \sum_{n=0}^{N-1}h'(nT_s)e^{-j2\pi\alpha nT_s}
\\\nonumber
&\approx \frac{E[|b|^{2}]}{N}\sum_{n=0}^{N-1}H'(\alpha_i)e^{-j2\pi\Delta\alpha nT_s} \\&= \frac{E[|b|^{2}] H'(\alpha_i)}{N} \frac{sin(\pi\Delta\alpha NT_s)}{sin(\pi\Delta\alpha T_s)}e^{-j\pi\Delta\alpha(N-1)T_s},
\label{eq:exp_R_si_2}
\end{align}
where $\Delta\alpha = |\alpha_t - \alpha_i|$ and $H'(\alpha_i)$ is Fourier series coefficient at $\alpha_i=1/T_h$ defined as $H(\alpha_i)=\frac{1}{T_h}\int_{0}^{T_h}h'(t)e^{j2\pi\alpha_i t}$. In (\ref{eq:exp_R_si_2}), the power of interference caused $s_i$ at $\alpha_t$ is negligible if   $\left(\frac{sin(\pi\Delta\alpha NT_s)}{sin(\pi\Delta\alpha T_s)}\right)^2 \approx 0$. In order to find a lower bound on number of samples $N$, we consider $\left(\frac{sin(\pi\Delta\alpha NT_s)}{sin(\pi\Delta\alpha T_s)}\right)^2 \approx 0$ if $\left(\frac{sin(\pi\Delta\alpha NT_s)}{sin(\pi\Delta\alpha T_s)}\right)^2 < 0.001$, i.e., the power of side-lobe is $< -30$ dB. This condition is satisfied if there are at least ten side-lobes of the sinc function $\left(\frac{sin(\pi\Delta\alpha NT_s)}{sin(\pi\Delta\alpha T_s)}\right)^2$ in the interval $[\alpha_t, \alpha_i]$. Therefore, the required lower bound on $N$ is $N>10 \lceil\frac{f_s}{\Delta \alpha}\rceil =10 \lceil \frac{f_s}{|\alpha_i-\alpha_t|} \rceil$.

\section{Proof: $\phi_k$ is a strictly monotonically decreasing function of $\rho_k$ and its value lies between 0 and 1.}
From the definition, $\phi_k = \frac{{\operatorname{var} ({{\hat R}_{{r_k}}^{\alpha_t}})}}
{{E[|{{\hat R}_{{r_k}}^{\alpha_t}}{|^2}]}}$, where ${\operatorname{var} ({{\hat R}_{{r_k}}^{\alpha_t}})}$ and ${E[|{{\hat R}_{{r_k}}^{\alpha_t}}{|^2}]}$ are given as
\begin{align}
\nonumber
&{\operatorname{var} ({{\hat R}_{{r_k}}^{\alpha_t}})} = p_{t,k}^2{\operatorname{var} ({{\hat R}_{{s_t}}^{\alpha_t}})} + p_{i,k}^2{\operatorname{var} ({{\hat R}_{{s_i}}^{\alpha_t}})} + {\operatorname{var} ({{\hat R}_{{w}}^{\alpha_t}})} 
\\\nonumber&+ p_{t,k}p_{i,k}E[|{{\hat R}_{{s_ts_i}}^{\alpha_t}}|^2] + p_{t,k}E[|{{\hat R}_{{s_tw}}^{\alpha_t}}|^2] + p_{i,k}E[|{{\hat R}_{{s_iw}}^{\alpha_t}}|^2],
\end{align}
\begin{align}
\nonumber
&{E[|{{\hat R}_{{r_k}}^{\alpha_t}}{|^2}]} = p_{t,k}^2 E[|{{\hat R}_{{s_t}}^{\alpha_t}}|^2]+ p_{i,k}^2E[|{{\hat R}_{{s_i}}^{\alpha_t}}|^2] + E[|{{\hat R}_{{w}}^{\alpha_t}}|^2] 
\\\nonumber &+ p_{t,k}p_{i,k}E[|{{\hat R}_{{s_ts_i}}^{\alpha_t}}|^2] + p_{t,k}E[|{{\hat R}_{{s_tw}}^{\alpha_t}}|^2] + p_{i,k}E[|{{\hat R}_{{s_iw}}^{\alpha_t}}|^2].
\end{align}
The above equations follows from $E[{{\hat R}_{{s_t}}^{\alpha_t}}{{\hat R}_{{s_ts_i}}^{\alpha_t*}}] = E[{{\hat R}_{{s_t}}^{\alpha_t}}{{\hat R}_{{s_tw}}^{\alpha_t*}}]=E[{{\hat R}_{{s_i}}^{\alpha_t}}{{\hat R}_{{s_ts_i}}^{\alpha_t*}}] = E[{{\hat R}_{{s_i}}^{\alpha_t}}{{\hat R}_{{s_tw}}^{\alpha_t*}}]=E[{{\hat R}_{{s_ts_i}}^{\alpha_t}}]=0$, which has been proved in the Appendix B. 
From Appendix B, $E[|{{\hat R}_{{w}}^{\alpha_t}}|^2]$  and ${\operatorname{var} ({{\hat R}_{{w}}^{\alpha_t}})}$ are proportional to $\frac{\sigma_w^4}{N^2}$  and $E[|{{\hat R}_{{s_tw}}^{\alpha_t}}|^2] = E[|{{\hat R}_{{s_iw}}^{\alpha_t}}|^2]  \approx \frac{2\sigma_w^2}{N}$. We assume that the terms $E[|{{\hat R}_{{w}}^{\alpha_t}}|^2]$, ${\operatorname{var} ({{\hat R}_{{w}}^{\alpha_t}})}$, $E[|{{\hat R}_{{s_tw}}^{\alpha_t}}|^2]$ and $ E[|{{\hat R}_{{s_iw}}^{\alpha_t}}|^2]$ are negligibly small for noise PSD = $-174$dBm/Hz, observed bandwidth at each CR = $f_s/2$ and $N > 10 \lceil f_s/\Delta \alpha \rceil$. For $N > 10 \lceil f_s/\Delta \alpha \rceil$, we have ${\operatorname{var} ({{\hat R}_{{s_i}}^{\alpha_t}})} =E[|{{\hat R}_{{s_i}}^{\alpha_t}}|^2] $, since $E[{{\hat R}_{{s_i}}^{\alpha_t}}] = 0$ as shown in Appendix C. Then, $\phi_k$ reduces to the following
\begin{align}
{\phi _k} &= \frac{{\operatorname{var} ({{\hat R}_{{r_k}}^{\alpha_t}})}}
{{E[|{{\hat R}_{{r_k}}^{\alpha_t}}{|^2}]}} = \frac{\rho_k^2\operatorname{var} ({{\hat R}_{{s_t}}^{\alpha_t}}) + E[|{{\hat R}_{{s_i}}^{\alpha_t}}{|^2}] + \rho_k E[|{{\hat R}_{{s_t}{s_i}}^{\alpha_t}}{|^2}]}
{\rho_k^2E[|{{\hat R}_{{s_t}}^{\alpha_t}}{|^2}] + E[|{{\hat R}_{{s_i}}^{\alpha_t}}{|^2}] + \rho_k E[|{{\hat R}_{{s_t}{s_i}}^{\alpha_t}}{|^2}]}
\nonumber
\\&= \frac{v_k}
{e_k} = \frac{\rho_k^2 v_{t} + e_{i} + \rho_k e_{ti}}
{\rho_k^2 e_t + e_i + \rho_k e_{ti}}.
\label{eq:app_phi2}
\end{align}
For simplicity of notations, operators ${\operatorname{var}(.)}$ and $E[|.|^2]$ are replaced by variables $v$ and $e$, respectively. The theoretical value of $\phi_k$ in (\ref{eq:app_phi2}) is computed using moments derived in Appendix B. We note that all the variables in above equation are non-negative. 

Let $\rho_{k1} > \rho_{k2}$, then we have ${\phi _{k1}} = \frac{\rho_{k1}^2 v_{t} + e_{i} + \rho_{k1} e_{ti}}
{\rho_{k1}^2 e_t + e_i + \rho_{k1} e_{ti}}
$ and ${\phi _{k2}} = \frac{\rho_{k2}^2 v_{t} + e_{i} + \rho_{k2} e_{ti}}
{\rho_{k2}^2 e_t + e_i + \rho_{k2} e_{ti}}$.
To prove that $\phi_{k1}<\phi_{k2}$, consider the following steps:
\begin{align}       
{\phi _{k1}} &\lesseqgtr {\phi _{k2}} 
\nonumber\\
\frac{{\rho _{k1}^2{v_t} + {v_i} + {\rho _{k1}}{v_{ti}}}}
{{\rho _{k1}^2{e_t} + {v_i} + {\rho _{k1}}{v_{ti}}}} &\lesseqgtr \frac{{\rho _{k2}^2{v_t} + {v_i} + {\rho _{k2}}{v_{ti}}}}
{{\rho _{k2}^2{e_t} + {v_i} + {\rho _{k2}}{v_{ti}}}}.
\end{align}
After cross-multiplying with denominators and removing the common terms on both sides, we have:
\begin{align}
 [{e_i}(\rho _{k1}^2 - \rho _{k2}^2) + {e_{ti}}{\rho _{k1}}{\rho _{k2}}({\rho _{k1}} - {\rho _{k2}})]({v_t} - {e_t}) &\lesseqgtr 0.
\label{eq:inequality}
\end{align}
Since $\rho_{k1}>\rho_{k2}$ and $v_t - e_t = \operatorname{var} ({{\hat R}_{{s_t}}}) - E[|{{\hat R}_{{s_t}}}{|^2}] = - |E[{\hat R}_{{s_t}}]|^2 < 0$, the LHS in the above equation is less than zero. Therefore, ${\phi _{k1}} < {\phi _{k2}}$. Hence, $\phi_k$ is a strictly monotonically decreasing function of $\rho_k$. Further, $\phi_k$ is a ratio of two non-negative quantities, $\phi_k \geq 0$. And,
\begin{align}
{\phi _k} &= \frac{{\operatorname{var} ({{\hat R}_{{r_k}}^{\alpha_t}})}}
{{E[|{{\hat R}_{{r_k}}^{\alpha_t}}{|^2}]}} = 1 - \frac{{|E[{{\hat R}_{{r_k}}^{\alpha_t}}{}]|^2}}{{E[|{{\hat R}_{{r_k}}^{\alpha_t}}{|^2}]}}.
\end{align}
Since $\frac{{|E[{{\hat R}_{{r_k}}^{\alpha_t}}{}]|^2}}{{E[|{{\hat R}_{{r_k}}^{\alpha_t}}{|^2}]}} \geq 0$, $\phi_k$ can not exceed 1. Therefore, $0 \leq \phi_k \leq 1$.
\section{Derivations required to compute $\mu_{v_s} $, $\sigma_{v_s}^2$, $\mu_{e_s} $, $\sigma_{e_s}^2$ and $\sigma_{v_se_s}$}
From Section \ref{sec:phi_k_M}, we have
$\mu_{v_{s}} = \operatorname{var}(\hat{R}_{r_k}^{\alpha_t})  \text{ and }
\sigma_{v_{s}}^2 = \frac{1}{M}\left[{\mu_4 - \frac{M-3}{M-1}\mu_{v_{s}}^2}\right]$,
where $\mu_4 = E\{|\hat{R}_{r_k}^{\alpha_t} - E[\hat{R}_{r_k}^{\alpha_t}]|^4\} = 2[\operatorname{var}({\hat{R}_{r_k}^{\alpha_t}})]^2 $ from \cite{Miller1969}. Further we have, $\mu_{e_{s}} =  E[|\hat{R}_{r_k}^{\alpha_t}|^2] \text{ and }
\sigma_{e_{s}}^2 = \operatorname{var}(|\hat{R}_{r_k}^{\alpha_t}|^2)/M
$.
Note that $|\hat{R}_{r_k}^{\alpha_t}|^2$ can be expressed in terms of $\boldsymbol{\hat{\theta}}$, $\mathbf{p_k}$ using (\ref{eq:vector_def}) and (\ref{eq:cac_rk_expanded}) as 
\begin{align}
|\hat{R}_{r_k}^{\alpha_t}|^2 = {{\left[ {{\boldsymbol{\hat \theta_r }^T}{\text{ }}{\boldsymbol{\hat \theta_i }^T}} \right]\left[ {\begin{array}{*{20}{c}}
   \mathbf{p_kp_k}^T & \boldsymbol{0}  \\
   \boldsymbol{0} & \mathbf{p_kp_k}^T  \\
 \end{array} } \right]\left[ \begin{gathered}
  {\boldsymbol{\hat \theta_r }} \hfill \\
  {\boldsymbol{\hat \theta_i }} \hfill \\ 
\end{gathered}  \right]}}
= {{{\boldsymbol{\hat \theta }^T}\mathbf{A_k}\boldsymbol{\hat \theta} }},
\end{align}
where $\mathbf{A_k} = \diag(\mathbf{p_kp_k}^T, \mathbf{p_kp_k}^T)$. Therefore, $|\hat{R}_{r_k}|^2$ is a quadratic form in Gaussian vector ${\boldsymbol{\hat \theta }}$. Hence, the mean and variance of $|\hat{R}_{r_k}|^2$ can be written using \cite[Eqn. 6]{Magnus1986} as follows:
\begin{align}
\nonumber
E[|\hat{R}_{r_k}^{\alpha_t}|^2] &= Tr(\mathbf{A_k}\boldsymbol{\Sigma_{\hat \theta}}) + E[\boldsymbol{\hat \theta}]^T \mathbf{A_k}E[\boldsymbol{\hat \theta}], \text{ }
\\
\operatorname{var}(|\hat{R}_{r_k}^{\alpha_t}|^2) &=  2Tr[(\mathbf{A_k} \boldsymbol{\Sigma_{\hat \theta}})^2] + 4E[\boldsymbol{\hat \theta}]^T \mathbf{A_k}\boldsymbol{\Sigma_{\hat \theta}} \mathbf{A_k}E[\boldsymbol{\hat \theta}]
\label{eq:app_e_3}
\end{align}
Computations of $E[\boldsymbol{\hat \theta}]$ and $\Sigma_{\hat \theta}$ are shown in Appendix A. $E[\hat{R}_{r_k}^{\alpha_t}]$ is also obtained from the moments derived in Appendix A. Using (\ref{eq:app_e_3}), $\mu_{v_s} $, $\sigma_{v_s}^2$, $\mu_{e_s} $, $\sigma_{e_s}^2$ are computed.

Next, we have to compute $E[{v_s}{e_s}] $ in order to obtain $\sigma_{v_se_s} = E[{v_s}{e_s}] -  \mu_{v_s} \mu_{e_s}$, where $v_s$ and $e_s$ are sample variance and sample second order moment of $\hat{R}_{r_k}^{\alpha_t}$ using $M$ samples. The sample mean ($m_s$), the sample second order moment and sample variance of $\hat{R}_{r_k}^{\alpha_t}$ are defined in (\ref{eq:sample_fvc}). Therefore, $E[{v_s}{e_s}] $ becomes
\begin{align}
E[{v_s}{e_s}] =\frac{M-1}{M}\left[E[v_s^2] + E[v_s |m_s|^2]\right],
\label{eq:corr_v_e}
\end{align}
 where $E[v_s^2] = \sigma_{v_s}^2 + \mu_{v_s}^2$, and  $E[v_s|m_s|^2]$ is obtained as follows. Although counter-intuitive, it has been shown in \cite{Shanmugam2008}, that the correlation coefficient between the sample mean $m_s$ and the sample variance $v_s$ are is zero if the samples are taken from a symmetric distribution such as Gaussian distribution. Therefore, the correlation coefficient between $v_s$ and $m_s$ is zero, i.e., $v_s$ and $m_s$ are uncorrelated.
Since $v_s$ and $m_s$ are Gaussian variables, they are also independent due to uncorrelatedness. Therefore, $v_s$ and $|m_s|^2$ are also independent which gives us $E[v_s|m_s|^2] = E[v_s]E[|m_s|^2] = \mu_{v_s}E[|m_s|^2]$, where $E[|m_s|^2] = var(\hat{R}_{r_k})/M + |E[\hat{R}_{r_k}]|^2 $. Substituting in (\ref{eq:corr_v_e}) we get:
$E[{v_s}{e_s}] = \frac{M-1}{M} \left[\sigma_{v_s}^2 + \mu_{v_s}^2 +\mu_{v_s}\left(\frac{\operatorname{var}(\hat{R}_{r_k}^{\alpha_t})}{M} + |E[\hat{R}_{r_k}^{\alpha_t}]|^2 \right) \right].$



\bibliographystyle{IEEEtran}
\bibliography{references_cyclic_wcl}

\end{document}